\documentclass[aps,pra,preprint,groupedaddress]{revtex4-1}

\usepackage{amsmath}
\usepackage{graphicx}
\usepackage[capitalise]{cleveref}
\usepackage{subcaption}
\usepackage{appendix}

\allowdisplaybreaks

\begin{document}

\title{Harmonic frequency combs in quantum cascade lasers: time-domain and frequency-domain theory}

\author{Yongrui Wang}
\affiliation{Department of Physics and Astronomy, Texas A\&M
University, College Station, TX, 77843 USA}
\author{Alexey Belyanin}
\email{belyanin@tamu.edu}
\affiliation{Department of Physics and Astronomy, Texas A\&M
University, College Station, TX, 77843 USA}

\date{\today}

\begin{abstract}
    Harmonic frequency combs, in which the lasing modes are separated by a period of tens of free spectral ranges  from each other, have been recently discovered in quantum cascade lasers (QCLs). There is an ongoing debate how the harmonic combs can be formed and stable under continuous pumping. Here we reproduce the harmonic state of lasing in QCLs by space-time-domain numerical simulations and show that the corresponding optical wave has a frequency-modulated (FM) nature, with the fundamental beat note in the electron population being suppressed. To understand the physics behind the formation and stability of the harmonic state, we develop a frequency-domain linearized analytic theory which analyzes the stability of single mode lasing with respect to the harmonic state formation and the instabilities of the resulting harmonic state. Our analysis shows how the instability of a single lasing mode can result in the growth of harmonic side modes. The coupling between the two side modes leads to opposite gains for FM and amplitude-modulated (AM) waves, and the FM gain is enhanced as compared to the AM gain. The dependence of the sideband gain spectrum on group velocity dispersion is also studied. While the instability gain of a single lasing mode does not prove that the harmonic state can be self-starting, we analyze the sideband instability gain in the presence of three strong harmonic modes and demonstrate that the harmonic state can be stable and self-supported. 
\end{abstract}

\maketitle

\section{Introduction}

Optical frequency combs \cite{udem2002} with equidistant separation and locked phases between the comb lines which serve as a ruler in the frequency domain, have revolutionized spectroscopy and metrology. A standard method of producing frequency combs is based on mode-locked lasers generating ultrashort pulses, which operate mostly in the near-infrared and visible ranges. The mid-infrared (mid-IR) and terahertz (THz) spectral regions, where most chemical compounds have strong spectral fingerprints, hold enormous potential for frequency comb applications. Conventional methods of extending frequency combs towards longer wavelengths relied on the nonlinear frequency down-conversion and requires open-air optical setups and high-power optical pumps. 

Quantum cascade laser (QCLs) \cite{faist2013book} are compact semiconductor laser sources which cover most of the mid-IR and parts of the THz spectral regions. Unfortunately, ultrafast gain relaxation in QCLs effectively prohibits passive mode locking, except theoretical proposals based on self-induced transparency solitons  \cite{menyuk2009,talukder2009}. Active mode locking schemes are less convenient and produce pulses of limited peak power \cite{barbieri2011,wang2009,wojcik2013,wang2015,revin2016,bachmann2016}, although there are some recent advances in this direction: see \cite{hillbr2020}, where a large modulation depth was achieved by careful design of the active region.  Note also that in the shorter part of the mid-IR range between 2-4\,$\mu$m passively mode-locked Type-I diode lasers \cite{feng2018,feng2020} and interband cascade lasers \cite{bagheri2018,hillbrand2019} have been recently demonstrated.  

So, generation of ultrashort pulses in QCLs has proven to be an uphill battle. That is why the realization that strong resonant nonlinearity of the gain transition itself is enough to trigger the formation of frequency combs in QCLs was such a pleasant surprise \cite{yu2009,wojcik2011,wojcikJMO2011}. It turned out that the phase-sensitive part of four-wave mixing (FWM) interactions mediated by the gain transition is enough to couple frequency and phases of QCL cavity modes into an equidistant comb; no additional nonlinear element is needed \cite{wojcik2011,wojcikJMO2011}. These early works showed synchronisation of combs belonging to different lateral modes, which was not very practical. In \cite{hugi2012} a much broader comb belonging to a single transverse mode in a QCL with ultra-broad gain spectrum was demonstrated, which opened the floodgate of QCL frequency comb efforts in the mid-IR and later in the THz  \cite{burghoff2014,wienold2014}. Besides conventional QCLs with Fabry-Perot (FP) cavity, frequency combs have also been reported recently in QCLs with ring cavity \cite{piccardo2019,meng2020}. 

Recent theoretical studies \cite{khurgin2014,tzenov2016,kazakov2017,opack2019} confirmed that it is FWM mediated by the ultrafast gain transition which remains the main mechanism responsible for the comb formation. However, unlike the shorter-wavelength combs generated by ultrashort-pulse lasers, in QCLs the underlying periodic modulation is linked to a predominantly FM optical wave \cite{khurgin2014}. In general, when the gain recovery time is much smaller than the roundtrip time, the amplitude modulation in the optical wave is most likely chaotic, as there is no mechanism to lock the phase between the modes. One should expect that a multimode spectrum can only be a frequency comb when the optical wave is mainly frequency modulated. However, an ideal FM wave cannot be realized; the AM is always present to some extent. This is because  the gain recovery time is short but still finite; see \cite{piccardoOptica2018} for a detailed discussion and also the recent work \cite{opack2019} where the effect of laser parameters on generating FM-dominated frequency combs was discussed. 

A new twist in the QCL frequency comb saga started a few years ago, when the harmonic state in QCL operation was discovered \cite{mansuripur2016}. In the harmonic state the lasing modes are separated in frequency domain by a large number of free spectral ranges (FSRs), often tens of FSRs, which makes it distinct from a conventional laser state, where laser modes are separated by one FSR. It was later verified that harmonic state is also a frequency comb, i.e.~the cavity modes are strictly equidistant and phase locked \cite{kazakov2017}. The separation between neighboring lasing modes in the harmonic combs reaches hundreds of GHz and even THz frequencies, which implies coherent (sub)picosecond intracavity modulation. Moreover, the separation can be tuned by optical seeding  \cite{piccardoAPL2018,piccardoOE2018}. Therefore, harmonic QCL combs provide direct link between mid-IR lasing and coherent (sub)THz generation and modulation \cite{piccardoOptica2018,piccardoPNAS2019,piccardoIEEE2019}, which makes them promising as (sub)THz coherent sources and transceivers. At the same time, such a large frequency interval between modes makes experimental characterization of harmonic combs very difficult and requires sophisticated experimental techniques \cite{piccardoOptica2018,piccardoPNAS2019,piccardoIEEE2019,piccardoPRL2019}. Some properties of the harmonic state still cannot be measured directly. Therefore, many aspects of the underlying physical mechanisms which give rise to complete suppression of neighboring cavity modes and the harmonic comb formation remain a mystery. This makes theoretical studies and in particular analytic theory insight into the harmonic comb physics particularly valuable. 

It is natural to start theoretical analysis with considering the instability of a QCL operating in a CW single mode regime. The instability of a single-mode QCL has been previously studied in \cite{gordon2008}, where it was shown that the instability of a FP cavity laser originates from the phase variation (FM optical wave) due to the spatial hole burning (SHB), whereas the Riesken-Nummedal-Graham-Haken (RNGH) instability \cite{risken1968,graham1968}  corresponding to an AM optical wave has a very high threshold, which cannot be reached in usual QCL operation. A later study \cite{mansuripur2016} showed that the instability corresponding to AM optical wave occurs at a large frequency detuning, and hence the development of a harmonic state could be explained. However, the authors neglected the fact that the perturbation of the population grating can experience diffusion, so the conclusions of that work were only valid when there is no diffusion. In another paper on the subject \cite{vukovic2017}, it was suggested that mid-IR QCLs with normal cavity length (a few mm long) have a low RNGH instability threshold, and it is even possible to have self pulsation in short cavity QCLs. The stability of QCLs and the formation of harmonic combs has been an active research area. 

In this work, we first use space- and time-domain numerical simulation in a QCL model which includes current injection, transport, and realistic boundary conditions to find the harmonic frequency comb regime and identify the phases of the spectral modes, as well as the space-time shape of the wave form, which are not directly observable in the experiment. To get a better understanding of how the harmonic state can be reached in QCLs, we develop analytic linearized theory of the sideband generation in the presence of a single strong mode and three harmonic strong modes. When one strong central mode is present, the instability gain is highest for side modes with phases corresponding to the FM wave. The peak gain of the FM instability is located at a frequency which is typically several or many FSRs away from the central mode, which indicates the tendency to develop the harmonic state, although does not prove its stability.  This part is an extension to our previous work \cite{kazakov2017}, where we mainly analyzed the frequency pulling due to the gain medium. To analyze the stability of the harmonic state,  we study the sideband instability in the presence of a three-mode harmonic state. We show that for certain parameters, the sideband instability gain of a three-mode harmonic state can be all negative, thanks to the coupling of a weak sideband mode to other modes far away from strong laser modes. This calculation gives some insight into why the harmonic state can be self-sustained and stable.

In \cref{{sec:numerical_simulation}}, we show the results of the numerical simulation of the harmonic state. In \cref{sec:linear_1mode}, we present the analysis of the stability of a single-mode lasing and in \cref{sec:linear_3mode}, we analyze the stability of a three-mode harmonic state.

\section{Numerical simulation of QCL harmonic combs}
\label{sec:numerical_simulation}

To reproduce the experimentally observed harmonic frequency combs and study their properties, we conduct space- and time-domain simulations of the QCL dynamics. The QCL model used is described in detail in our previous paper \cite{wang2015}, except that here we assume that the same constant DC bias is applied along the whole cavity length. 

In this model, the QCL active region includes the injector subband, and the upper and lower laser subbands. The injector in one period also serves as the collector subband in the previous period. Besides the optical transition, we include the non-radiative relaxations between the subbands. For the electron transport from the injector state to the upper laser state, we use Kazarinov-Suris formula \cite{kazarinov1972} to calculate the resonant tunneling current. In this approach, the bias on the QCL acts as an input parameter, whereas the injection current is calculated along with other output variables for a given laser state. Spontaneous fluctuations of the field and polarization are included as delta-correlated random noise sources. 

In the experiment \cite{mansuripur2016}, it is shown that the laser at a given pumping level can have different operation states, very long timescale instabilities, and even hysteresis, in which the laser regime at a given injection current depends on whether it was reached by increasing or decreasing the current.  This indicates that a harmonic state in a QCL is only locally stable, and one should not expect that the laser can evolve into the harmonic state from arbitrary initial conditions. In our simulations, we find that the laser may only predictably reach a harmonic state with a low-power seeding at a side mode frequency, similarly to the experiment in \cite{piccardoAPL2018}.

\begin{figure}[htb]
	\includegraphics[width=0.5\textwidth]{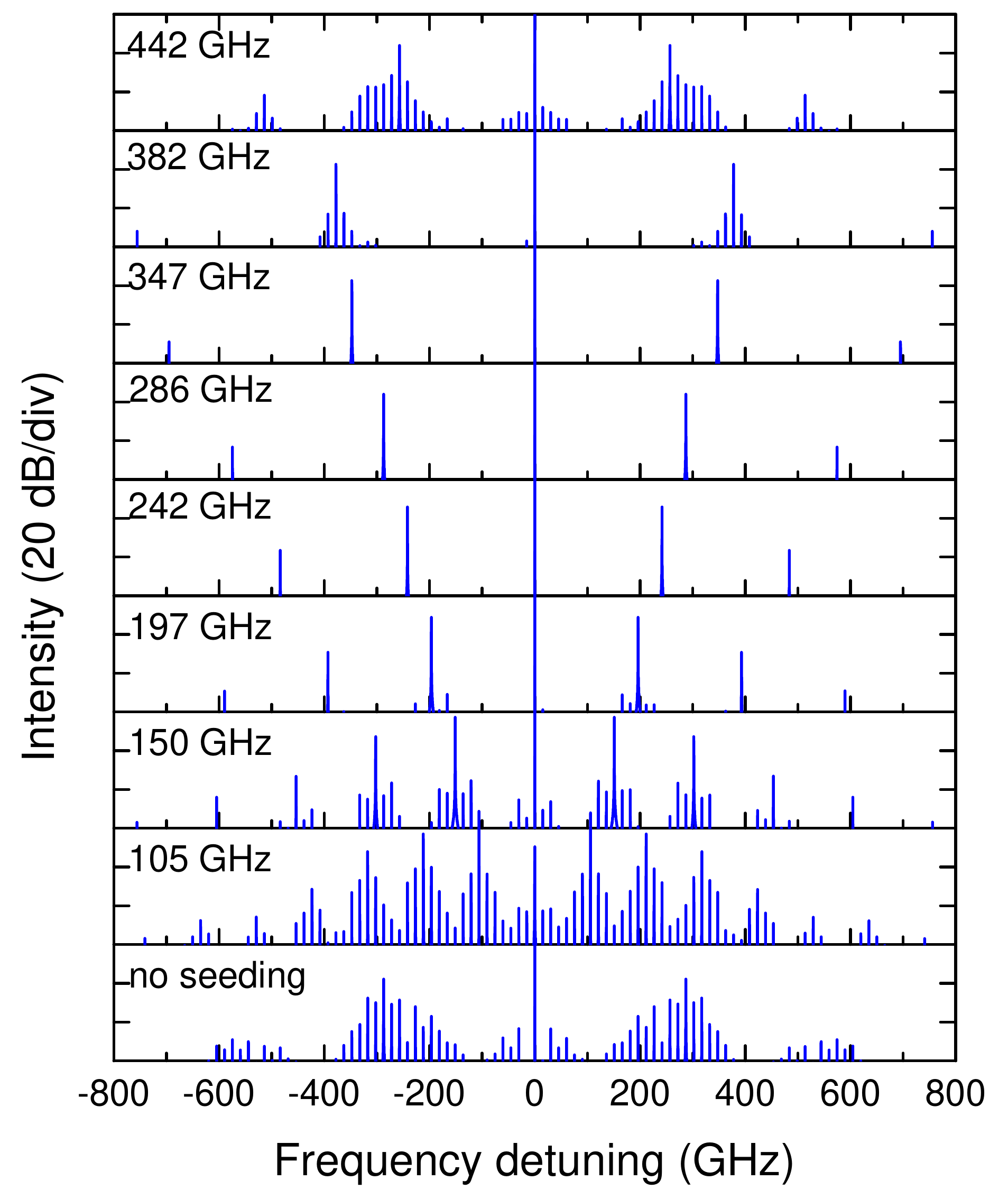}
	\caption{Dependence of laser spectrum on detuning of a weak seeding signal. The seeding signal has an amplitude of 0.001 kV/cm, and the pumping is fixed at $p=1.08$. The laser cavity length is 3 mm, which corresponds to the FSR of 15.2 GHz.  \label{Fig:spectrum_dw_dependence}}
\end{figure}

With a set of parameters corresponding to the gain relaxation time $T_1 = 0.5\,\mathrm{ps}$, homogeneous broadening time $T_2=0.02\,\mathrm{ps}$, reflection coefficients at both laser facets equal to 0.5, and pumping levels corresponding to experiment, we are able to obtain the harmonic states by preparing the initial state as a weak central mode, with $E_0 \sim 0.01\,\mathrm{kV/cm}$, and applying very weak seeding signal $E_s = 0.001\,\mathrm{kV/cm}$ at variable frequency detunings. In comparison, a typical electric field of the laser in our simulation is of the order of 10 kV/cm at the currents about 10\% above threshold. In \cref{Fig:spectrum_dw_dependence}, we show the spectra of the laser output for different seeding frequencies, with pumping level fixed at $p=J/J_{th}=1.08$. The laser cavity length is 3 mm, which corresponds to the FSR of 15.2 GHz. 

We run each simulation for 4000 roundtrips ($\mathrm{T_{round}}$), and the weak seeding is turned on for the first 3000 roundtrips, and then turned off. In all of the simulations, the laser output is observed to be stable for the last 1000 roundtrips. The spectrum is calculated using the time sequence of the output field for the last 300 roundtrips. The results show that harmonic states can be achieved for a particular range of side-mode detunings compatible with experiment. Even though seeding is used to initiate the formation of the harmonic states, they are sustained after the seeding is turned off and therefore are compatible with QCL dynamics in the presence of a strong laser field in the cavity.  So, the obtained results show that the harmonic states can be reached in numerical simulations, and one can study their properties  from the simulation results.

In the lowest panel of \cref{Fig:spectrum_dw_dependence}, the spectrum is obtained when the weak seeding is not applied. This spectrum consists of the central mode, and also bumps of modes on both sides. The appearance of the bumps in the spectrum can be interpreted by the gain spectrum we calculated in the linear analysis, where we find that the net gain is positive for a range of frequency detunings, see Sec. \ref{sec:linear_1mode}. If the seeding frequency resides in the range of the bump, we can generally obtain a harmonic state, such as the cases when the seeding frequencies are 242, 286, and 347 GHz. Otherwise we mainly get the mixed harmonic state, as defined in \cite{mansuripur2016}, where the harmonic modes are surrounded by bumps of neighboring modes, see, e.g.~the spectra when the seeding frequency is 150 GHz or 442 GHz.

\begin{figure}[htb]
	\includegraphics[width=0.5\textwidth]{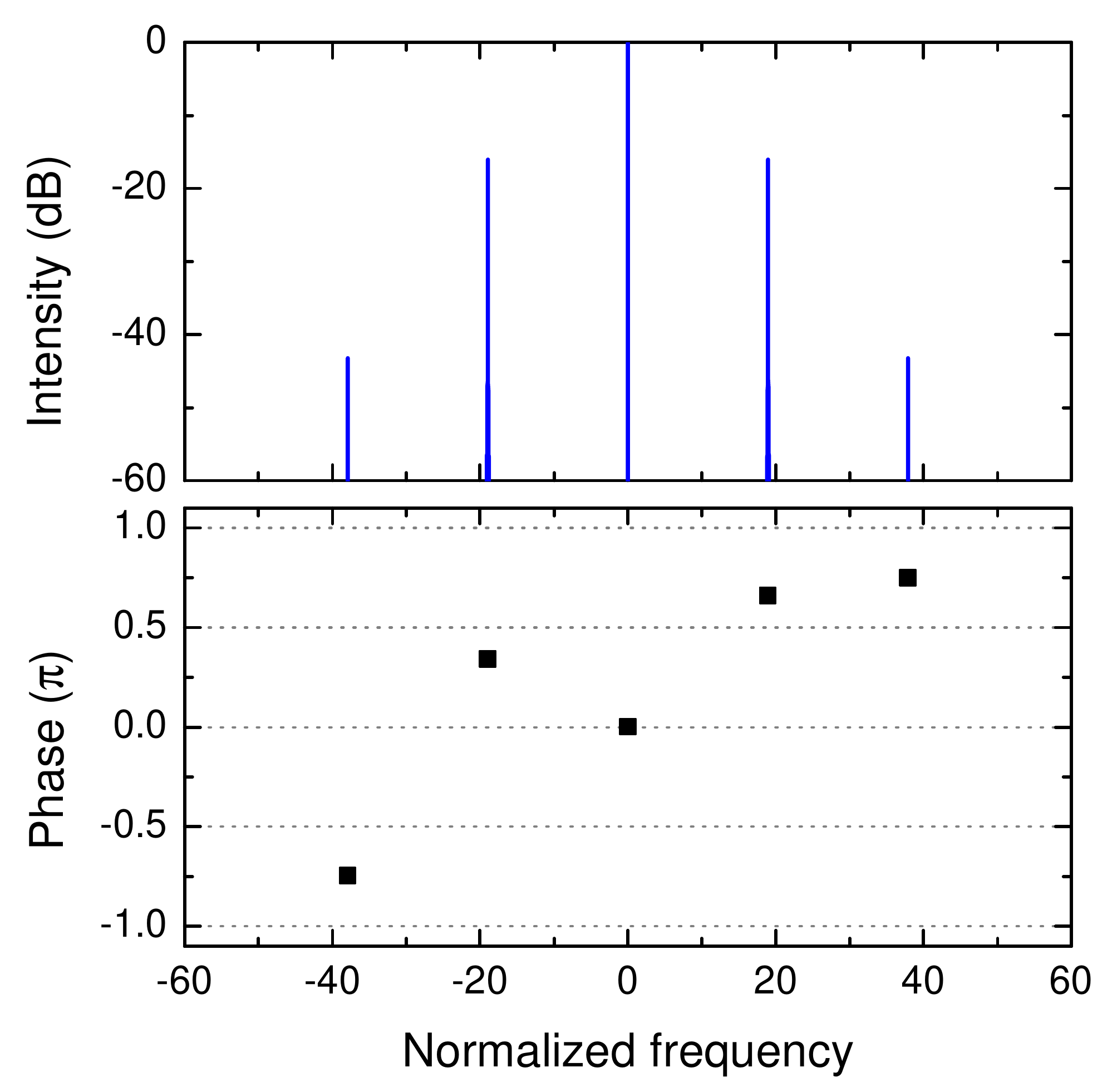}
	\caption{The amplitudes and phases of the lasing modes when side-mode detuning is 286 GHz. The phases (labeled as $\phi_0$, $\phi_+$, $\phi_-$, $\phi_{2+}$, and $\phi_{2-}$) satisfy the relations $(\phi_+-\phi_0)-(\phi_0-\phi_-) = \pi$ and $(\phi_{2+}-\phi_0)-(\phi_0-\phi_{2-}) = 0$.  \label{Fig:spectrum_phase}}
\end{figure}

To learn more about the properties of the harmonic states, we first try to get the phase relation between the laser modes. In \cref{Fig:spectrum_phase}, we show the amplitude and phase of the spectrum of a harmonic state, in which the seeding frequency is detuned from the central mode by 286\,GHz, and pumping level is at $p=1.08$. The phases of the central mode, the first two side modes, and the second two side modes are denoted as $\phi_0$, $\phi_\pm$, and $\phi_{2\pm}$, respectively, where the $+$ and $-$ indicate positive and negative frequency detunings. One can see that the phases of the side modes at the fundamental frequency are close to the relation $(\phi_+-\phi_0)-(\phi_0-\phi_-) = \pi$, which indicates that the time-variation of the output field is frequency-modulated (FM) at the fundamental frequency \cite{mansuripur2016}. On the other hand, the phases of the side modes at the second harmonic satisfy $(\phi_{2+}-\phi_0)-(\phi_0-\phi_{2-}) \approx 0$. To understand this relation, note that the beat note at the fundamental frequency originates mainly from two couplings, 0/$+$ and 0/$-$, which cancel each other, as seen in their phase relation. However, there are also two weak contributions to the beat note at the fundamental frequency, $+$/$2+$ and $-$/$2-$. Their contribution is canceled too, so that there is a phase relation $(\phi_{2+}-\phi_+)-(\phi_--\phi_{2-}) = \pi$. Then we can obtain the relation $(\phi_{2+}-\phi_0)-(\phi_0-\phi_{2-}) = 0$. The beat note at the second harmonic comes from three couplings, 0/$2+$, 0/$2-$ and $+$/$-$. According to the intensities of the modes shown in \cref{Fig:spectrum_phase}, the main contribution comes from the coupling $+$/$-$. This beat note cannot be canceled out, so the output optical field is amplitude-modulated (AM) at the second harmonic. 

As was argued in \cite{piccardoPRL2019}, the QCL with an ultrafast gain relaxation time tries to avoid all the beat notes in order to maximize the output power. In the language of laser dynamics, coupled oscillations of the intracavity intensity and population inversion that are slow as compared to the gain recovery time cannot be supported by the laser. However, the gain recovery time $T_1 = 0.5$ ps is short but finite and moreover, of the same order of magnitude as the timescales of higher harmonics of beatnotes between lasing modes in the harmonic state. That is why the AM is observed at the second harmonic. Mathematically, one can see that with only a few modes in the harmonic state the laser cannot be exactly frequency-modulated, so only the beat note at the fundamental frequency is canceled out \cite{thanktoby}.

\begin{figure}[htb]
	\includegraphics[width=0.5\textwidth]{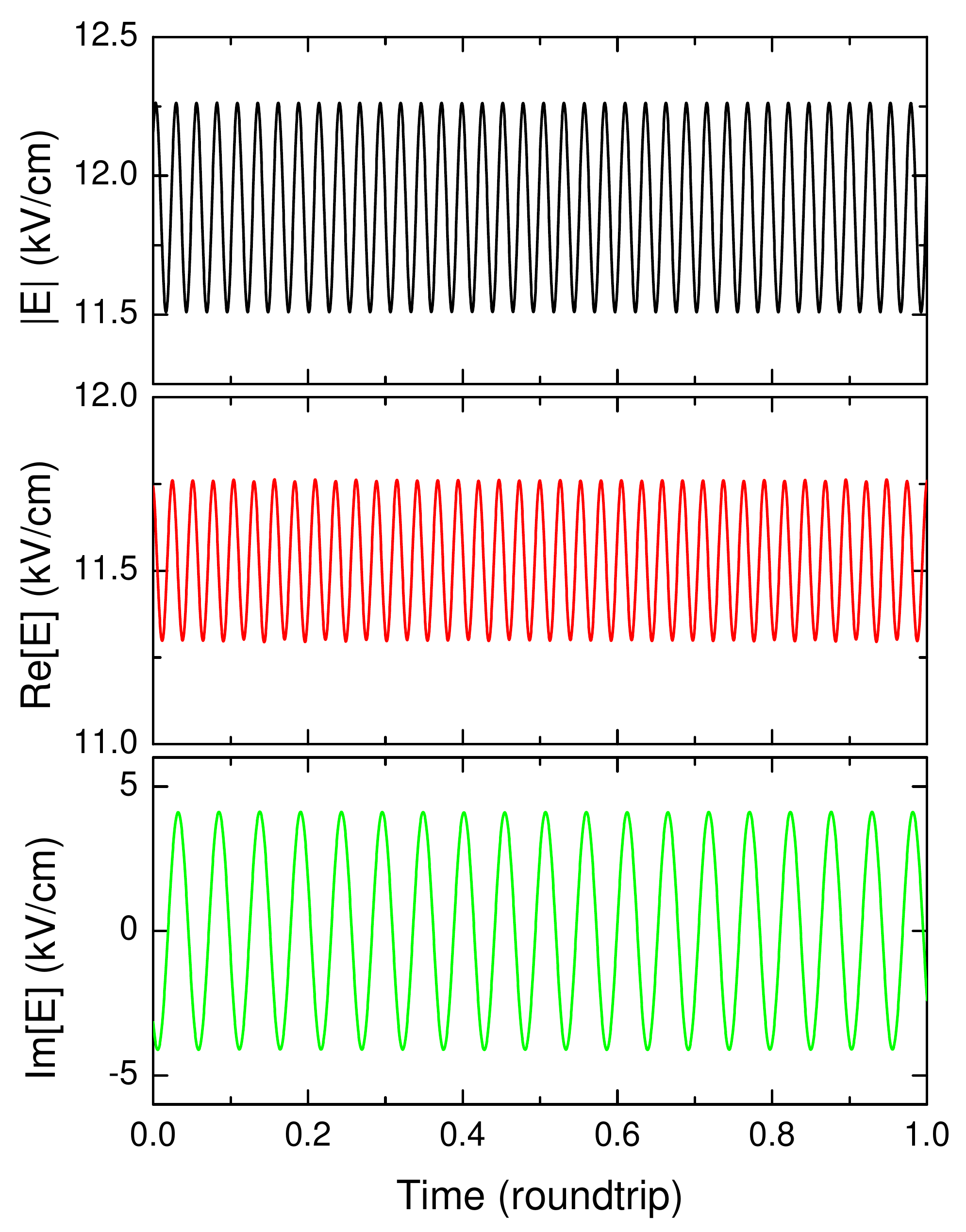}
	\caption{The time evolution of the laser field at a fixed point in the cavity when the side-mode detuning is 286\,GHz. It shows clearly the FM at the fundamental frequency and AM at the second harmonic.  \label{Fig:wave_form}}
\end{figure}

The FM and AM features can be clearly seen in the plots of a waveform shown in \cref{Fig:wave_form}, where the oscillation of the amplitude is at the frequency of the second harmonic, while the oscillation of the imaginary part of the field is at the fundamental frequency. 


\begin{figure}[htb]
	\includegraphics[width=0.5\textwidth]{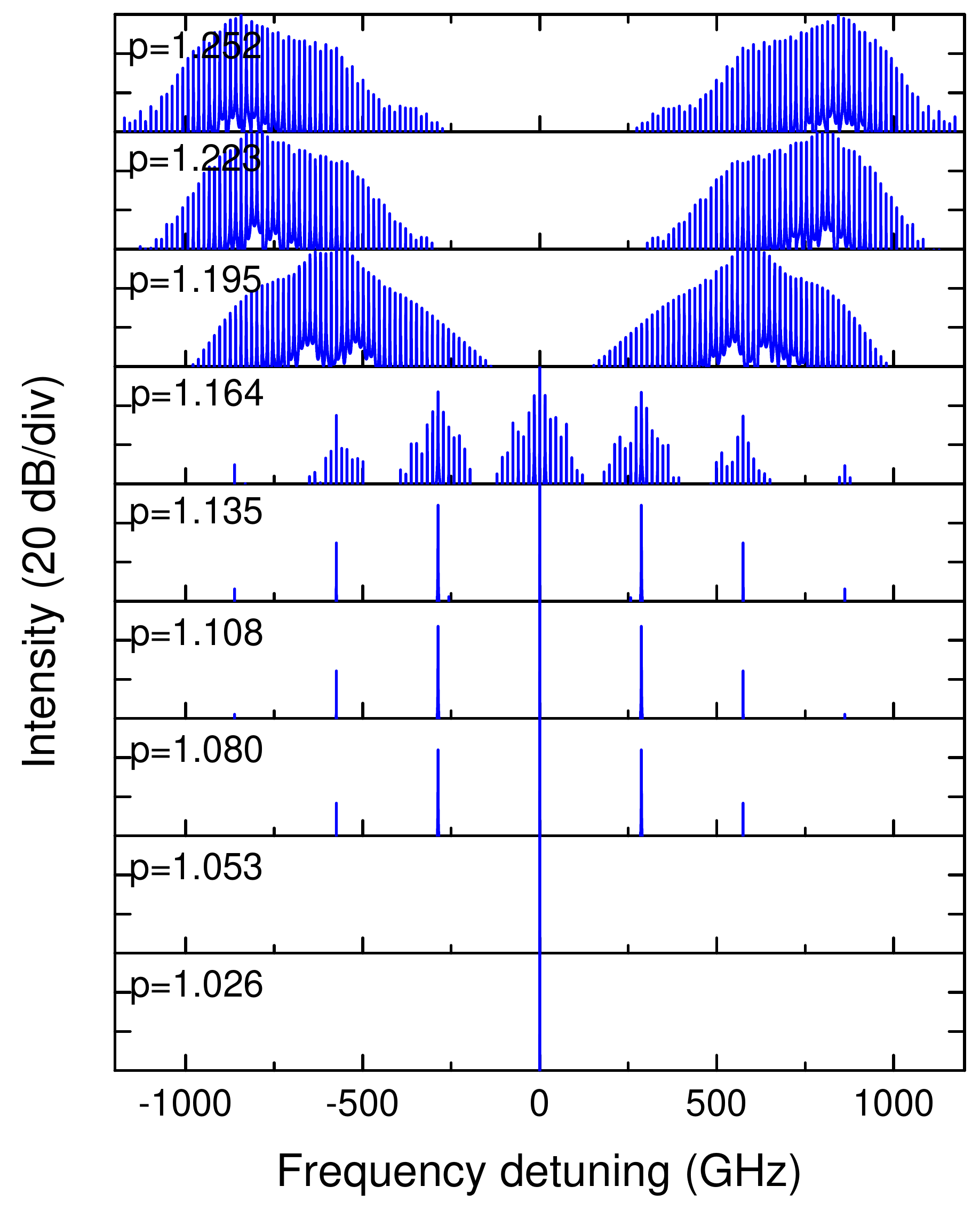}
	\caption{Dependence of spectrum on pumping level. The seeding is weak, with an amplitude of 0.001 kV/cm, and frequency detuning of 286\,GHz.  \label{Fig:spectrum_pumping_dependence}}
\end{figure}

Now we fix the frequency detuning of the weak seeding to be $f_s = 286$ GHz, and change the pumping rate. The resulting spectra are shown in \cref{Fig:spectrum_pumping_dependence}. The results indicate that stable harmonic state can be reached in a finite range of injection currents. For low pumping levels, as in the lowest two panels in \cref{Fig:spectrum_pumping_dependence}, the central mode is not strong enough to maintain the side modes with visible intensity. On the other hand, when the pumping level is high, the harmonic state becomes unstable because the side modes cannot extract all the gain from the central mode, so other modes will be amplified too, which results in a mixed-harmonic state or a dense state with all cavity mode lasing. 

The harmonic state we obtained in the numerical simulation is found to be stable for a few thousands of roundtrips, as checked by visually watching the output waveform. To investigate the stability for much longer simulation time, we conducted a simulation for 20,000 roundtrips. The result shows that the laser exhibits long-term instabilities, in which harmonic state deteriorates and eventually evolves into dense state; see the supplemental video [Supplemental video link] for the long-term evolution of the spectrum. The evolution of the spectrum is generally very slow, with a time scale of hundreds of ns, which is much longer than any dynamic timescales in QCLs. The underlying physics behind this slow evolution is not clear at this moment. To get more insight into the origins, physics, and stability of the harmonic state, we developed linearized analytic theory to follow the evolution of small perturbations to single mode and harmonic state lasing, as described in the next two sections.

\section{Instability of a single-mode lasing and formation of a harmonic state}
\label{sec:linear_1mode}
A QCL, as any other laser, starts lasing in a single mode just above threshold. This is seen in our numerical simulations \cref{Fig:spectrum_pumping_dependence}, and this is of course what the experiments show. As the pumping is increased, the formation of the harmonic state can be naturally explained if the instability gain of a single-mode state peaks at frequencies multiple FSRs away from the first lasing mode. With this objective in mind, we analyze the instability of a single-mode state by obtaining analytic solution to the linearized equations governing small perturbations of a single-mode lasing.

\subsection{The model}
Since the goal is to obtain a simple enough analytic model, we use a two-level active region model to describe resonant coupling of a gain transition to cavity modes. The Maxwell-Bloch equations which describe the interaction between the electrons and the electric field of laser modes are given by
\begin{align}
\label{Eq:MB}
&\partial_t \rho_{ul} = - \left( i \omega_{ul} + 
\gamma_2 \right) \rho_{ul} - i \frac{d {\cal E}}{\hbar} \Delta ,  \nonumber \\
&\partial_t \Delta = - \gamma_1 (\Delta-\Delta_p) - 2 i \frac{d {\cal E}}{\hbar} (\rho_{ul} - \rho_{ul}^\ast) + D \frac{\partial^2\Delta}{\partial z^2} , \\
&\partial_z^2 {\cal E} - \frac{\sigma}{\epsilon_0 c^2} \partial_t {\cal E} - \frac{n^2}{c^2} \partial_t^2 {\cal E} = \kappa d \partial_t^2 (\rho_{ul} + \rho_{ul}^\ast) ,  \nonumber 
\end{align}
where the subscripts $u,l$ mark upper and lower laser states and $\rho_{ul}$ is the off-diagonal element of the density matrix, or the optical coherence, which gives rise to the polarization that acts as a source in the wave equation. Furthermore, $\Delta$ is the population inversion, ${\cal E}$ is the optical electric field, $d$ is the dipole matrix element, $\gamma_1 = 1/T_1$ is the gain relaxation rate, $\gamma_2 =1/T_2$ is the dephasing rate, $\omega_{ul}$ is the transition frequencyy, $n$ is the refractive index, $\sigma$ is the conductivity which accounts for the waveguide loss, $D$ is the diffusion coefficient of the population inversion grating, and $\kappa = \Gamma / \epsilon_0 c^2 L_p$, with $\Gamma$ the modal overlap factor and $L_p$ the period of the active region. The pumping is described by $\Delta_p$. Throughout this manuscript, when numerical calculation is conducted, $\omega_{ul}$ corresponds to a wavelength of 6.2 $\mu$m, $d/e$ = 2 nm, and $D= 46$ cm$^2$/s \cite{wang2009}. 

In order to get analytic results, we restrict the analysis to a QCL with high reflectivity facets. We assume that the reflections on both facets have zero phase shift and choose the coordinates so that the laser cavity is between $z = 0$ and $z = L$. Then the spatial variation of a mode has the form of $\cos(k z)$, with $k L = \mathrm{integer}\times\pi$. 

For a mode with frequency $\omega$ and wave vector $k$, the optical field and coherence can be written as
\begin{align}
\label{Eq:amplitudes_MW}
&{\cal E} = E \cos(k z) e^{-i \omega t} + \mathrm{c.c.} ,  \nonumber \\
&\rho_{ul} = \eta \cos(k z) e^{-i \omega t} , 
\end{align}
where the envelopes $E$ and $\eta$ are slowly varying functions of time. Substituting them into the Maxwell wave equation, we obtain
\begin{align}
\label{Eq:MWeq}
&\phantom{{}={}} 2 i \frac{ n^2(\omega) \omega}{c^2} \partial_t E + i \frac{\sigma(\omega) \omega}{\epsilon_0 c^2} E + \left\{ - \frac{\sigma(\omega)}{\epsilon_0 c^2} \partial_t E - \frac{ n^2(\omega)}{c^2} \partial_t^2 E + \left( \frac{n^2(\omega)\omega^2}{c^2} - k^2 \right) E \right\}  \nonumber \\
&= - \kappa d \omega^2 \eta + \left\{ \kappa d \left( - 2 i \omega \partial_t \eta + \partial_t^2 \eta \right) \right\} .
\end{align}
The terms in the two curly brackets are generally small in magnitude compared to other terms, and they determine the dispersion relation. We may drop the $\partial_t$ terms there. Also, we assume that the frequency $\omega$ is close to the transition frequency $\omega_{ul}$, and replace all the $\omega$ by $\omega_{ul}$ except in the leading term in the dispersion. Then the Maxwell wave equation becomes
\begin{align}
2 i \frac{ n^2(\omega_{ul}) \omega_{ul}}{c^2} \partial_t E + i \frac{\sigma(\omega_{ul}) \omega_{ul}}{\epsilon_0 c^2} E + 2 \frac{n(\omega_{ul})\omega_{ul}}{c}  \left( \frac{n(\omega)\omega}{c} - k \right) E   
= - \kappa d \omega_{ul}^2 \eta  .
\end{align}
In a more compact form, it can be written as
\begin{align}
\label{Eq:MWeq_simplified}
\partial_t E = - l_t E + i \frac{c}{n(\omega_{ul})}  \left( \frac{n(\omega)\omega}{c} - k \right) E  + i \frac{\kappa d \omega_{ul} c^2}{2 n^2(\omega_{ul})} \eta  ,
\end{align}
where $l_t = \sigma(\omega_{ul})/(2 n^2(\omega_{ul})\epsilon_0)$ is the loss in time domain, which is related to the propagation loss $l$ by $l = l_t n(\omega_{ul})/c$. 


\subsection{Single-mode solution}	
When there is only one lasing mode in the laser cavity, the optical field, coherence, and population inversion can be written as
\begin{align}
\label{Eq:amplitudes_1mode}
{\cal E} &= E_0 \cos(k_0 z) e^{-i \omega_0 t} + \mathrm{c.c.} ,  \nonumber \\
\rho_{ul} &= \eta_0 \cos(k_0 z) e^{-i \omega_0 t} ,  \nonumber \\
\Delta &= \Delta_0 + \Delta_2 \cos(2 k_0 z)  ,
\end{align}
where we have included the population grating to the order of $|E_0|^2$. In principle, the polarization can have higher-order terms $\propto \cos(3k_0 z)$, $\cos(5k_0 z)$, $\cdots$, and $\Delta$ can have higher-order terms $\propto \cos(4k_0 z)$, $\cos(6k_0 z)$, $\cdots$. Including these terms would result in the optical susceptibility $\chi_0=\eta_0/E_0$ containing terms beyond the order of $|E_0|^2$. When the Rabi frequency $d |E_0|/\hbar$ is much smaller than $\sqrt{\gamma_1 \gamma_2}$, the contribution of the polarization grating and other higher-order terms is a small perturbation which can be neglected. If this condition does not hold, there is no clear cut where to truncate the higher-order terms because they are not converging. So, we will assume the condition $d |E_0|/\hbar < \sqrt{\gamma_1 \gamma_2}$ to hold in the analysis and always keep the optical susceptibility to the order of $|E_0|^2$. 

  
Using the ansatz of \cref{Eq:amplitudes_1mode}, the equations for the envelopes of coherence and population become
\begin{align}
\label{Eq:polarization_1mode}
&\partial_t \eta_0 = \left[ i (\omega_0-\omega_{ul}) - \gamma_2 \right] \eta_0  - i \frac{d}{\hbar} \left( E_0 \Delta_0 + \frac{1}{2} E_0 \Delta_2 \right)  ,  \nonumber \\
&\partial_t \Delta_0 = - \gamma_1 (\Delta_0-\Delta_p) - i \frac{d}{\hbar} \left(  E_0^\ast \eta_0 - E_0 \eta_0^\ast \right) , \nonumber \\
&\partial_t \Delta_2 = - \gamma_g \Delta_2 - i \frac{d}{\hbar} \left( E_0^\ast \eta_0 - E_0 \eta_0^\ast \right) , 
\end{align}
where $\gamma_g = \gamma_1+4 k_0^2 D$. We also define $T_g = \gamma_g^{-1}$, which will be used later. 

When the pumping is increased, lasing starts at the gain peak $\omega = \omega_{ul}$. The threshold population inversion is 
\begin{equation}
\Delta_{th} = \frac{2 n(\omega_{ul}) \hbar \gamma_2}{\kappa d^2 c \omega_{ul}} l .
\end{equation}
We assume that the single-mode frequency is at the gain peak $\omega_0 = \omega_{ul}$, as this is the most likely scenario. Defining the pumping level by $p = \Delta_p / \Delta_{th}$, the steady-state solution is 
\begin{align}
\label{Eq:sol_1mode}
&k_0 = \frac{n(\omega_0)\omega_0}{c} , \nonumber \\
&\left( \frac{d|E_0|}{\hbar} \right)^2 = (p-1) \frac{1}{(2 T_1 + T_g)T_2} , \nonumber \\
&\eta_0 = - i \frac{d T_2 \Delta_{th}}{\hbar} E_0, \nonumber \\
&\Delta_0 = \left[1 + T_g T_2 \left( \frac{d|E_0|}{\hbar} \right)^2 \right] \Delta_{th} , \nonumber  \\
&\Delta_2 = - 2 T_g T_2 \left( \frac{d|E_0|}{\hbar} \right)^2 \Delta_{th} .
\end{align}
As stated before, we have kept the susceptibility to the order of $|E_0|^2$, or $(p-1)$, as the two are proportional to each other. This also requires the populations to be kept to the $|E_0|^2$ order. The population grating $\Delta_2$ is negative, and its effect is to decrease the gain of the lasing mode, which is natural because population grating is generated by the optical field, so there must be a negative feedback to the field itself to make it stable. To maintain a zero net gain, the population $\Delta_0$ is above the threshold value. In such a situation, if a weak mode does not couple to $\Delta_2$ and has a frequency close to $\omega_0$, it should experience positive net gain, and hence may be amplified. However, such a weak mode can couple to the lasing mode via FWM interaction, which leads to new sideband generation and modifies the resulting sideband gain. This will be analyzed in the next section.

\subsection{Single-mode stability}	
\label{sec:1mode_stability}
With the single-mode state known, we would like to analyze its stability by calculating the linear gain of other weak modes in its presence. Due to the existence of the strong central mode, a weak side mode can generate another weak mode detuned symmetrically  with respect to the strong mode via FWM process. To make this analysis more general, we will work with a central mode at arbitrary frequency. Denote it as mode $r$, with frequency $\omega_r$ and wave vector $k_r$. The side modes are denoted with subscripts $s\pm$, with frequency detunings $\pm d\omega$, and wave-vector detunings $\pm d k$. Then we can write the fields and density matrix elements as
\begin{align}
\label{Eq:amplitudes_case_1}
&{\cal E} = E_r \cos(k_r z) e^{-i \omega_r t} + E_{s+} \cos((k_r+ dk) z) e^{-i (\omega_r+d\omega) t} + E_{s-} \cos((k_r-dk) z) e^{-i (\omega_r-d\omega) t} \nonumber \\
&+ \mathrm{c.c.} ,  \nonumber \\
&\rho_{ul} = \eta_r \cos(k_r z) e^{-i \omega_r t} + \eta_{s+} \cos((k_r+dk) z) e^{-i (\omega_r+d\omega) t} + \eta_{s-} \cos((k_r-dk) z) e^{-i (\omega_r-d\omega) t}  ,  \nonumber \\
&\Delta = \Delta_0 + \left[ \Delta_{s+} \cos(dk z) e^{-id\omega t} + \mathrm{c.c.} \right]  \nonumber \\
&+ \Delta_{2r} \cos(2 k_r z) + \left[ \Delta_{2s+} \cos((2k_r + dk) z) e^{-id\omega t} + \Delta_{2s-} \cos((2k_r - dk) z) e^{-id\omega t} + \mathrm{c.c.} \right] ,
\end{align}
where $E_r$, $\eta_r$, $\Delta_0$, and $\Delta_{2r}$ are from the solution of the single-mode state. In this analysis, we assume that $d\omega$ and $dk$ are nonzero, namely the weak modes do not coincide with the central mode. When the weak mode has the same spatial profile as the central mode, there is no instability as the gain is anti-correlated with the magnitude of the central mode. In the expansion of population inversion $\Delta$, we include not only the usual population pulsation term which contains $\Delta_{s+}$, but also terms corresponding to the pulsation in the population grating, which contain $\Delta_{2s+}$ and $\Delta_{2s-}$. This is necessary as we will see later: all of these terms couple with $\eta_{s\pm}$, so they affect the time evolution of the fields in the two side modes.


With the ansatz given in \cref{Eq:amplitudes_case_1}, the dynamics of the coherences and populations of the weak modes is given by
\begin{align}
\label{Eq:eta_dynamics}
&\partial_t \eta_{s+} = \left( i (\omega_r+d\omega-\omega_{ul}) - \gamma_2 \right) \eta_{s+} 
- i \frac{d}{\hbar} \left( \frac{1}{2} E_r \Delta_{s+} + \frac{1}{2} E_r \Delta_{2s+} + E_{s+} \Delta_0  \right)  ,  \nonumber  \\
&\partial_t \eta_{s-} = \left( i (\omega_r-d\omega-\omega_{ul}) - \gamma_2 \right) \eta_{s-} 
- i \frac{d}{\hbar} \left( \frac{1}{2} E_r \Delta_{s+}^\ast + \frac{1}{2} E_r \Delta_{2s-}^\ast + E_{s-} \Delta_0  \right) ,
\end{align}
and
\begin{align}
\label{Eq:popu_dynamics}
&\partial_t \Delta_{s+} = \left( id\omega - \gamma_1 \right) \Delta_{s+} - i \frac{d}{\hbar} \left( E_r^\ast \eta_{s+} - E_r \eta_{s-}^\ast - E_{s+} \eta_r^\ast + E_{s-}^\ast \eta_r \right) , \nonumber \\
&\partial_t \Delta_{2s+} = \left( id\omega - \gamma_g \right) \Delta_{2s+} - i \frac{d}{\hbar} \left( E_r^\ast \eta_{s+} - E_{s+} \eta_r^\ast \right) ,  \nonumber \\
&\partial_t \Delta_{2s-} = \left( id\omega - \gamma_g \right) \Delta_{2s-} - i \frac{d}{\hbar} \left( - E_r \eta_{s-}^\ast + E_{s-}^\ast \eta_r \right)  .
\end{align}
Assuming that the time variation of the envelopes is slow as compared to $\gamma_1$, $\gamma_g$ and $\gamma_2$, we may take the time derivatives in the above equations to be zero, so the coherences can be expressed in terms of the fields of the weak modes as
\begin{align}
\label{Eq:pol_E_1mode}
&\eta_{s+} = \left[ \chi_0(\omega_r + d\omega) + \alpha^r_{++} \right] E_{s+} + \alpha^r_{+-} E_{s-}^\ast , \nonumber \\
&\eta_{s-} = \left[ \chi_0(\omega_r - d\omega) + \alpha^r_{--} \right] E_{s-} + \alpha^r_{-+} E_{s+}^\ast ,
\end{align}
where 
\begin{align}
\label{Eq:chi0_def}
\chi_0(\omega) = \frac{d}{\hbar} \frac{\Delta_0}{\omega-\omega_{ul}+i\gamma_2} ,
\end{align}
and to the order of $|E_r|^2$, the coefficients $\alpha$ are given by
\begin{align}
\label{Eq:alpha_def}
\alpha^r_{++} 
&= \frac{d}{2\hbar} \frac{1}{\omega_r+d\omega-\omega_{ul}+i\gamma_2} \left( \frac{1}{d\omega+i\gamma_1} + \frac{1}{d\omega+i\gamma_g} \right) \nonumber \\
& \times \left[ \Delta_{th} \left(\frac{d|E_r|}{\hbar}\right)^2 \frac{1}{\omega_r+d\omega-\omega_{ul}+i\gamma_2}  
- \frac{d}{\hbar} E_r \eta_r^\ast \right] , \nonumber \\
\alpha^r_{+-} 
&= \frac{d}{2\hbar} \frac{1}{(\omega_r+d\omega-\omega_{ul}+i\gamma_2) (d\omega+i\gamma_1)} \nonumber \\
&\times \left[ - \Delta_{th} \left(\frac{d E_r}{\hbar}\right)^2 \frac{1}{\omega_r-d\omega-\omega_{ul}-i\gamma_2}  + \frac{d}{\hbar} E_r \eta_r \right]  , \nonumber \\
\alpha^r_{--}
&= \alpha^r_{++}|_{d\omega \rightarrow -d\omega}  , \nonumber \\
\alpha^r_{-+}
&= \alpha^r_{+-}|_{d\omega \rightarrow -d\omega}  .
\end{align}
The coherences $\eta_{s\pm}$ contain the terms due to direct excitation by the field of the side mode with the same index,  represented by $\chi_0(\omega)+\alpha^r_{++(--)}$, and  the cross coupling between the two side modes and the central mode, represented by $\alpha^r_{+-(-+)}$. Using the single-mode solution given in \cref{Eq:sol_1mode}, where $\omega_r=\omega_{ul}$, the expressions for $\chi_0$ and $\alpha$'s are reduced to
\begin{align}
    \label{Eq:chi0_1mode}
    \chi_0(\omega_r \pm d\omega) = \frac{d \Delta_{th}}{\hbar} \left[ 1 + \left(\frac{d|E_0|}{\hbar}\right)^2 \frac{1}{\gamma_g\gamma_2} \right] \frac{1}{\pm d\omega+i\gamma_2} ,
\end{align}
and
\begin{align}
\label{Eq:alpha_1mode}
\alpha^r_{++} 
&= \frac{d \Delta_{th}}{2\hbar} \left(\frac{d|E_0|}{\hbar}\right)^2 \frac{1}{d\omega+i\gamma_2} \left( \frac{1}{d\omega+i\gamma_1} + \frac{1}{d\omega+i\gamma_g} \right)
 \left( \frac{1}{d\omega+i\gamma_2} + \frac{1}{i\gamma_2} \right)  
 , \nonumber \\
\alpha^r_{+-} 
&= \frac{d \Delta_{th}}{2\hbar} \left(\frac{d E_0}{\hbar}\right)^2 \frac{1}{(d\omega+i\gamma_2) (d\omega+i\gamma_1)}   \left( \frac{1}{d\omega+i\gamma_2} + \frac{1}{i\gamma_2} \right)    , \nonumber \\
\alpha^r_{--}
&= \alpha^r_{++}|_{d\omega \rightarrow -d\omega}  , \nonumber \\
\alpha^r_{-+}
&= \alpha^r_{+-}|_{d\omega \rightarrow -d\omega}  .
\end{align}

The coherences in \cref{Eq:pol_E_1mode} can be plugged into the the wave equations \cref{Eq:MWeq_simplified} for each side mode, to get the coupled equations for the two side modes which can be written in a matrix form as
\begin{align}
\partial_t \left[ E_s \right]
= \left[ B \right] \left[ E_s \right] ,
\end{align}
with
\begin{align}
\left[ E_s \right] =
\begin{bmatrix}
E_{s+} &  E_{s-}^{\ast}
\end{bmatrix}^\mathrm{T} ,
\end{align}
and
\begin{align}
\left[ B \right] = - l_t + \left[ B \right]_{\mathrm{phase}} + \left[ B \right]_{\mathrm{direct}} + \left[ B \right]_{\mathrm{cross}} ,
\end{align}
where the part of matrix $[B]$ corresponding to phase relations is given by
\begin{align}
\left[ B \right]_{\mathrm{phase}} 
= i \frac{c}{n(\omega_{ul})} 
\begin{bmatrix} 
\frac{n(\omega_{s+})\omega_{s+}}{c} - k_{s+} & 0 \\
0 & -\frac{n(\omega_{s-})\omega_{s-}}{c} + k_{s-}   
\end{bmatrix}  ,
\end{align}
the part of matrix $[B]$ corresponding to the direct coupling is given by
\begin{align}
\left[ B \right]_{\mathrm{direct}} 
=
i \frac{\kappa d \omega_{ul} c^2}{2 n^2(\omega_{ul})}
\begin{bmatrix} 
 \chi_0(\omega_{s+}) + \alpha_{++}^r & 0  \\
0 & -\chi_0^\ast(\omega_{s-}) - \alpha_{--}^{r\ast}
\end{bmatrix}  ,
\end{align}
and the part of matrix $[B]$ corresponding to the cross coupling is given by
\begin{align}
\left[ B \right]_{\mathrm{cross}} 
=
i \frac{\kappa d \omega_{ul} c^2}{2 n^2(\omega_{ul})}
\begin{bmatrix}
0  & \alpha_{+-}^r \\
-\alpha_{-+}^{r\ast} & 0 
\end{bmatrix} .
\end{align}
In the above equations $\omega_{s\pm}=\omega_r\pm d\omega$ and $k_{s\pm}=k_r\pm dk$. 

The gain of the side modes is determined by the eigenvalues of matrix $[B]$. If the real part of an eigenvalue is greater than 0, there is instability. The imaginary part of the eigenvalue can always be tuned to zero by adjusting the value of $dk$ accordingly, as the dependence of matrix $\left[ B \right]_{\mathrm{phase}}$ on $dk$ is linear with a coefficient proportional to the unit matrix. To see this, the $[B]_{\mathrm{phase}}$ can be written as
\begin{align}
&\phantom{{}={}} \left[ B \right]_{\mathrm{phase}} \left( i \frac{c}{n(\omega_{ul})} \right)^{-1}    \nonumber \\
&=
\begin{bmatrix} 
\frac{[n(\omega_{s+})-n(\omega_r)] \omega_{s+}}{c} & 0 \\
0 & -\frac{[n(\omega_{s-})-n(\omega_r)]\omega_{s-}}{c}   
\end{bmatrix}  \nonumber \\
&+ 
\left( \frac{n(\omega_r) \omega_r }{c} - k_r \right)
\begin{bmatrix} 
1 & 0 \\
0 & -1   
\end{bmatrix} 
+ 
\left( \frac{n(\omega_r) d \omega }{c} - dk \right)
\begin{bmatrix} 
1 & 0 \\
0 & 1   
\end{bmatrix} .
\end{align}
So, the term containing $dk$ is really proportional to the unit matrix. Also, we can account for group velocity dispersion (GVD) here. To do this, the refractive index can be written as $n(\omega) = n(\omega_r) + \frac{1}{2} c \beta_2 (\omega-\omega_r)$, where $\beta_2$ is the GVD coefficient. Then the phase matrix becomes
\begin{align}
&\phantom{{}={}} \left[ B \right]_{\mathrm{phase}}  \left( i \frac{c}{n(\omega_{ul})} \right)^{-1}   \nonumber \\
&=
\begin{bmatrix} 
\frac{1}{2} \beta_2 d\omega (\omega_r+d\omega) & 0 \\
0 & \frac{1}{2} \beta_2 d\omega (\omega_r-d\omega)   
\end{bmatrix}  
+ 
\left( \frac{n(\omega_r) \omega_r }{c} - k_r \right)
\begin{bmatrix} 
1 & 0 \\
0 & -1   
\end{bmatrix} 
+ 
\left( \frac{n(\omega_r) d \omega }{c} - dk \right)
\begin{bmatrix} 
1 & 0 \\
0 & 1   
\end{bmatrix}   \nonumber \\
&=
\left( \frac{1}{2} \beta_2 d\omega^2 + \frac{n(\omega_r) \omega_r }{c} - k_r \right)
\begin{bmatrix} 
1 & 0 \\
0 & -1   
\end{bmatrix} 
+ 
\left( \left( \frac{1}{2} \beta_2 \omega_r + \frac{n(\omega_r) }{c} \right) d\omega - dk \right)
\begin{bmatrix} 
1 & 0 \\
0 & 1   
\end{bmatrix} .
\end{align}
Only the first term can affect the eigenstate, as the second term is proportional to a unit matrix.

In the single-mode solution given in \cref{Eq:sol_1mode}, $\omega_0=\omega_{ul}$ is assumed. If we further assume $n(\omega)=n(\omega_{ul})$ for any frequency $\omega$, namely that there is no dispersion in the waveguide refractive index, then the eigenvectors of matrix $[B]$ satisfy $|E_{s+}|=|E_{s-}|$ and the phase $\phi = \phi(E_{s+})+\phi(E_{s-})-2\phi(E_{0})$ is equal to either 0 or $\pi$, because the matrix $[B]$ satisfies $B_{11}=B_{22}$ and $B_{12}=B_{21}$.

\begin{figure}[htb]
	\centering
	\begin{subfigure}{0.45\textwidth}
		\centering
		\includegraphics[width=\linewidth]{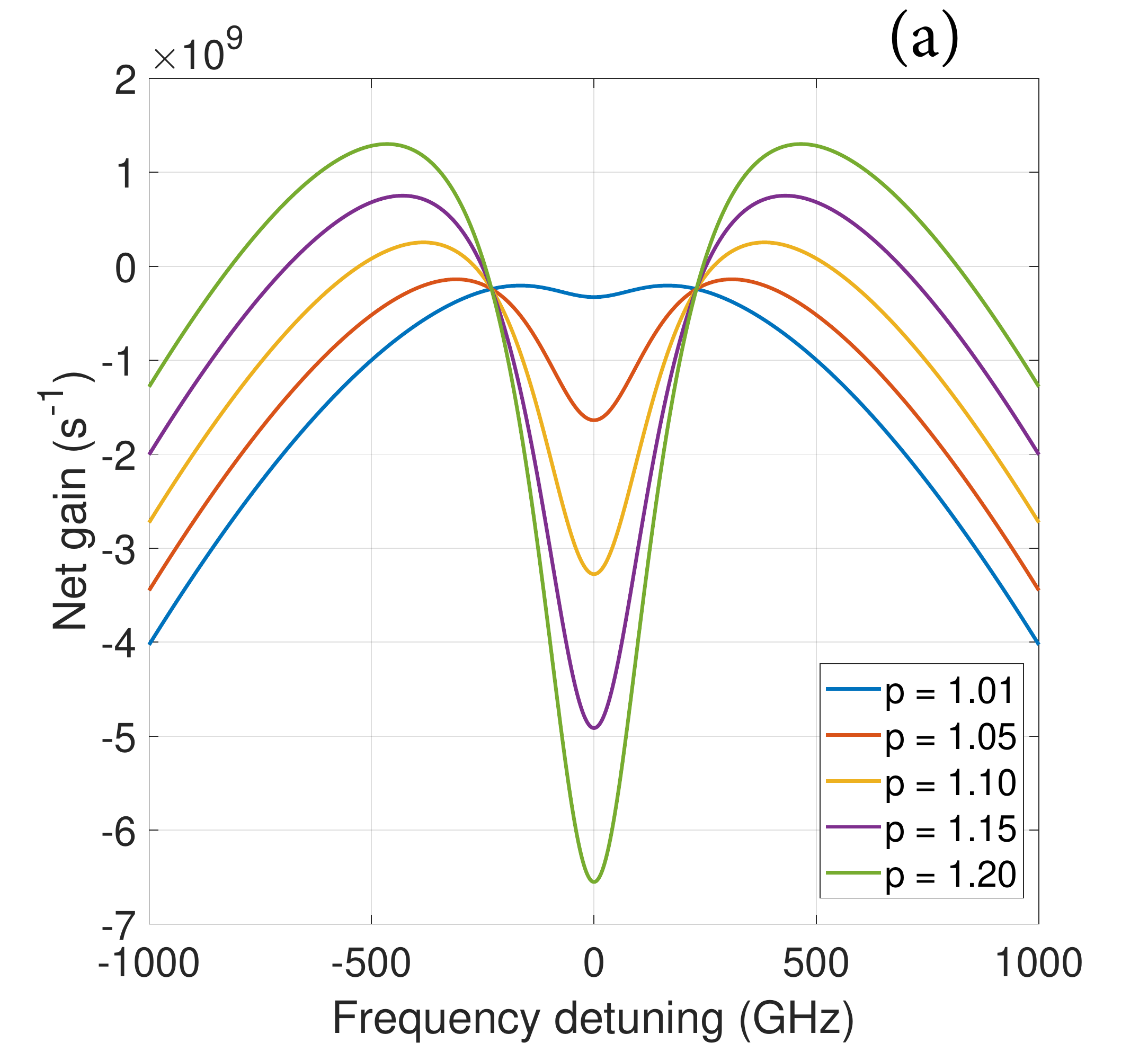}
	\end{subfigure}
	\begin{subfigure}{0.45\textwidth}
		\includegraphics[width=\linewidth]{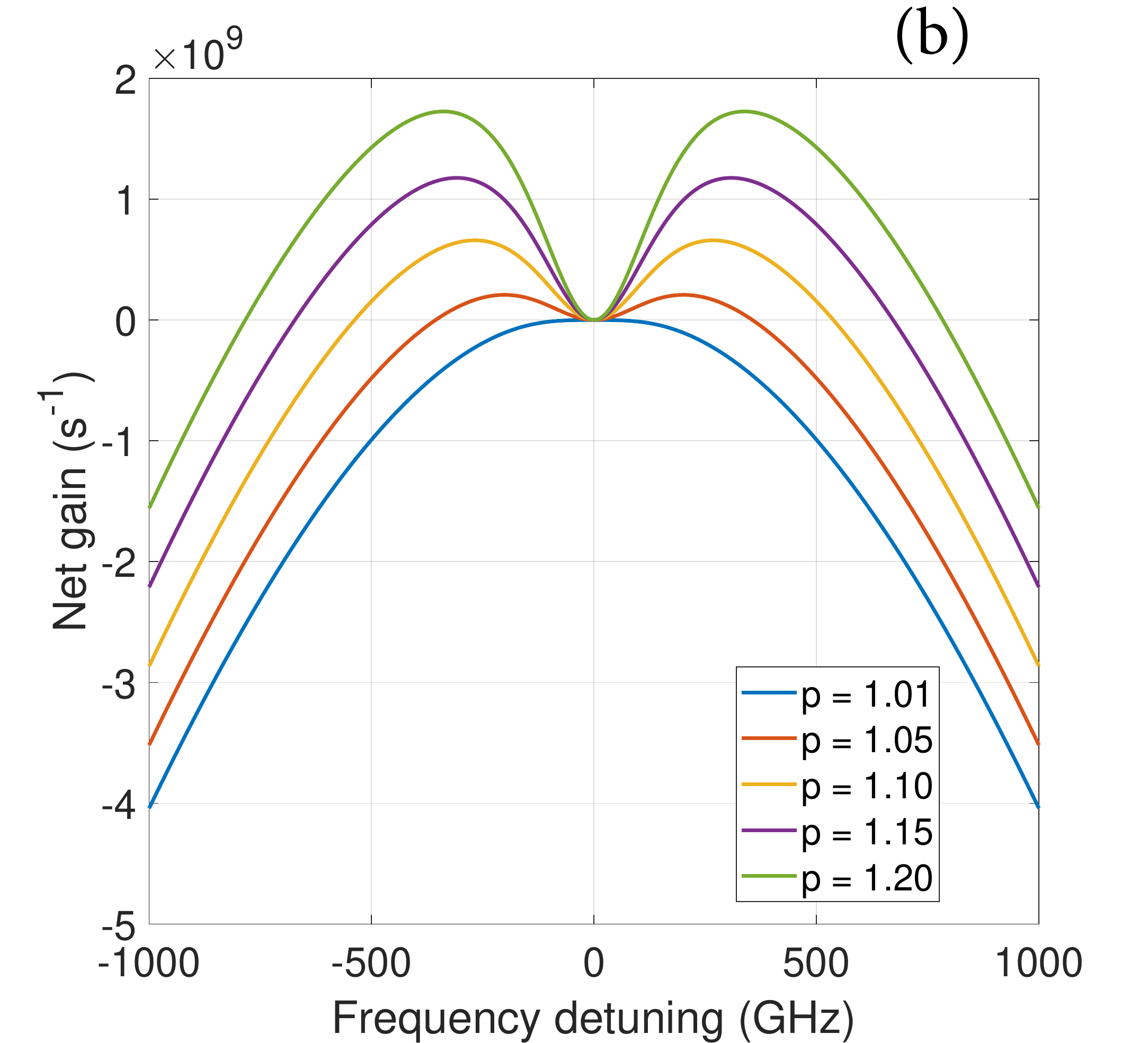}
	\end{subfigure}
	\caption{ Instability gain of the side modes for different pumping levels with the side modes having phase relations (a) $\phi = 0$, and (b) $\phi=\pi$. } .
	\label{Fig:gain_p_dep}
\end{figure}

In \cref{Fig:gain_p_dep}, we show the dependence of the gain spectrum on the pumping level, for $\phi=0$ and $\phi=\pi$. Here, $\phi=0$ and $\phi=\pi$ correspond to an AM and FM optical wave, respectively \cite{mansuripur2016}. For $\phi=\pi$, the maximum net gain is always positive starting at $p=1.05$. For $\phi=0$, the net gain is first negative and then positive when the pumping is increased. This AM instability has a much lower threshold than the RNGH instability analyzed in \cite{gordon2008}, indicating some discrepancy between the calculating methods. As the laser has both AM and FM instabilities starting at $p=1.10$, the nature of the state that the laser will evolve into at the nonlinear stage is uncertain. However, since the instability threshold is lower for FM-like side modes, one can expect that the laser first evolves into a multimode state corresponding to an FM optical wave. With the pumping further increased, the evolution of the laser will depend on the actual laser state it has reached, and cannot be determined by the linearized single-mode stability analysis.

To find the threshold of the instability, it is enough to consider the case of $\phi=\pi$. One can see that the instability always occurs at a frequency detuning close to zero, so the condition for the instability is determined by
\begin{align}
\frac{\partial}{\partial (d\omega)^2} g(d\omega)|_{\phi=\pi} > 0 ,
\end{align}
where $g(d\omega)$ is the net gain of the side modes. Using the analytical expression of the gain spectrum, we find the following instability condition: 
\begin{equation}
p-1 > \frac{2\gamma_g^2(\gamma_1+2\gamma_g)} { \gamma_1 (2\gamma_2^2+3\gamma_2\gamma_g+2\gamma_g^2) } . 
\end{equation} 
In mid-IR QCLs, the dephasing rate is usually much higher than the population relaxation rate, so we may assume $\gamma_2 \gg \gamma_1$. Also, if the  diffusion is not very strong, $\gamma_g \approx \gamma_1$ can be assumed, and the threshold condition becomes
\begin{equation}
p-1 > 3 \left( \frac{\gamma_1}{\gamma_2} \right)^2 .
\end{equation} 


\begin{figure}[htb]
	\centering
	\begin{subfigure}{0.45\textwidth}
		\centering
		\includegraphics[width=\linewidth]{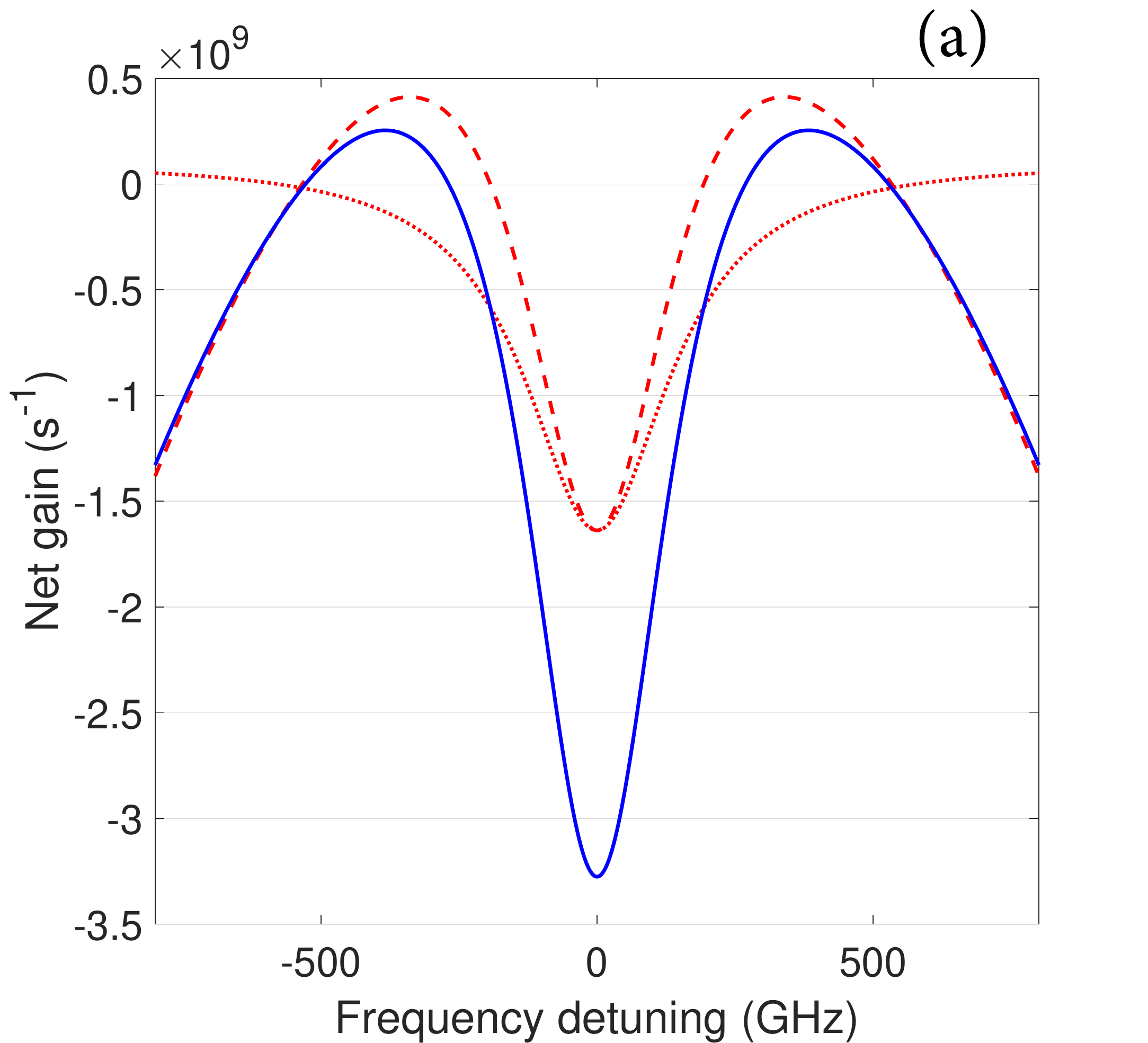}
	\end{subfigure}
	\begin{subfigure}{0.45\textwidth}
		\includegraphics[width=\linewidth]{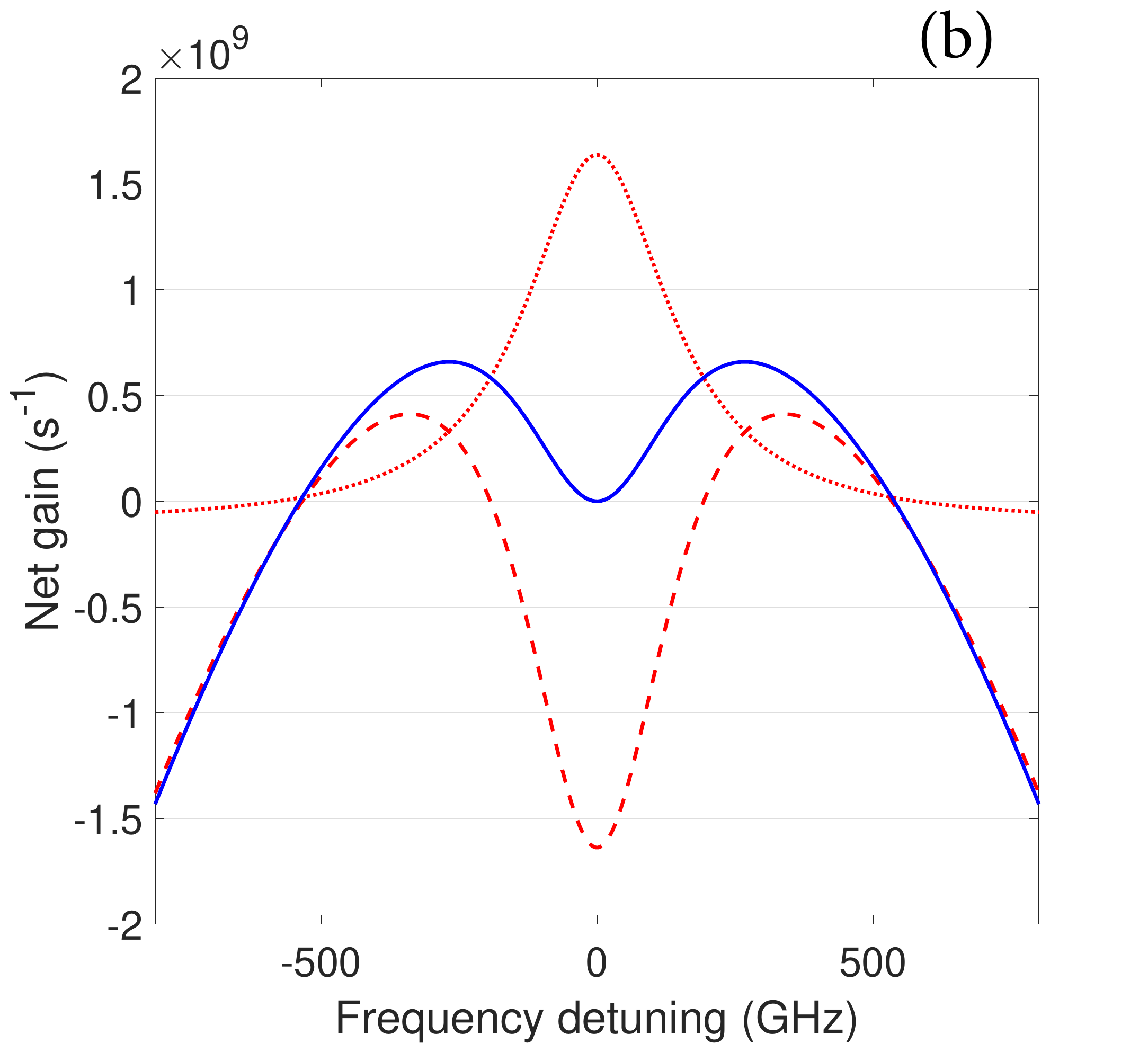}
	\end{subfigure}
	\caption{ The instability and contributions from the direct and cross couplings for (a) $\phi = 0$, and (b) $\phi=\pi$. Solid blue line, red dashed line, and red dotted line denote the total gain, direct gain, and cross gain, respectively. Pumping is at $p=1.1$. } .
	\label{Fig:gain_direct_and_cross}
\end{figure}

As the matrix [B] contains the direct and cross coupling terms, we can analyze their contributions to the sidemode gain separately. For the side mode $s+$, they are $g_{\mathrm{direct}}(E_{s+}) = \mathrm{Re}[ \left( [B]_{\mathrm{direct}} [E_s] \right)_{s+} / E_{s+} ]$, and $g_{\mathrm{cross}}(E_{s+}) = \mathrm{Re}[ \left( [B]_{\mathrm{cross}} [E_s] \right)_{s+} / E_{s+} ]$, with the eigenstate $[E_s] = [ E_{s+} ~  E_{s-}^{\ast} ]^{\mathrm{T}}$. The expressions for the gain of the symmetric side mode $s-$ are similar. In \cref{Fig:gain_direct_and_cross}, we show the instability gain at $p = 1.1$, together with separate contributions of the direct gain and cross gain. The direct gains for $\phi=0$ and $\phi=\pi$ are the same, while the cross gains have opposite sign for the two phases. The cross gain for $\phi=\pi$ is positive at low frequencies, so the instability gain at low frequencies is higher for $\phi=\pi$ than for $\phi=0$.

\subsection{The role of population pulsations}

\begin{figure}[htb]
	\centering
	\begin{subfigure}{0.45\textwidth}
		\centering
		\includegraphics[width=\linewidth]{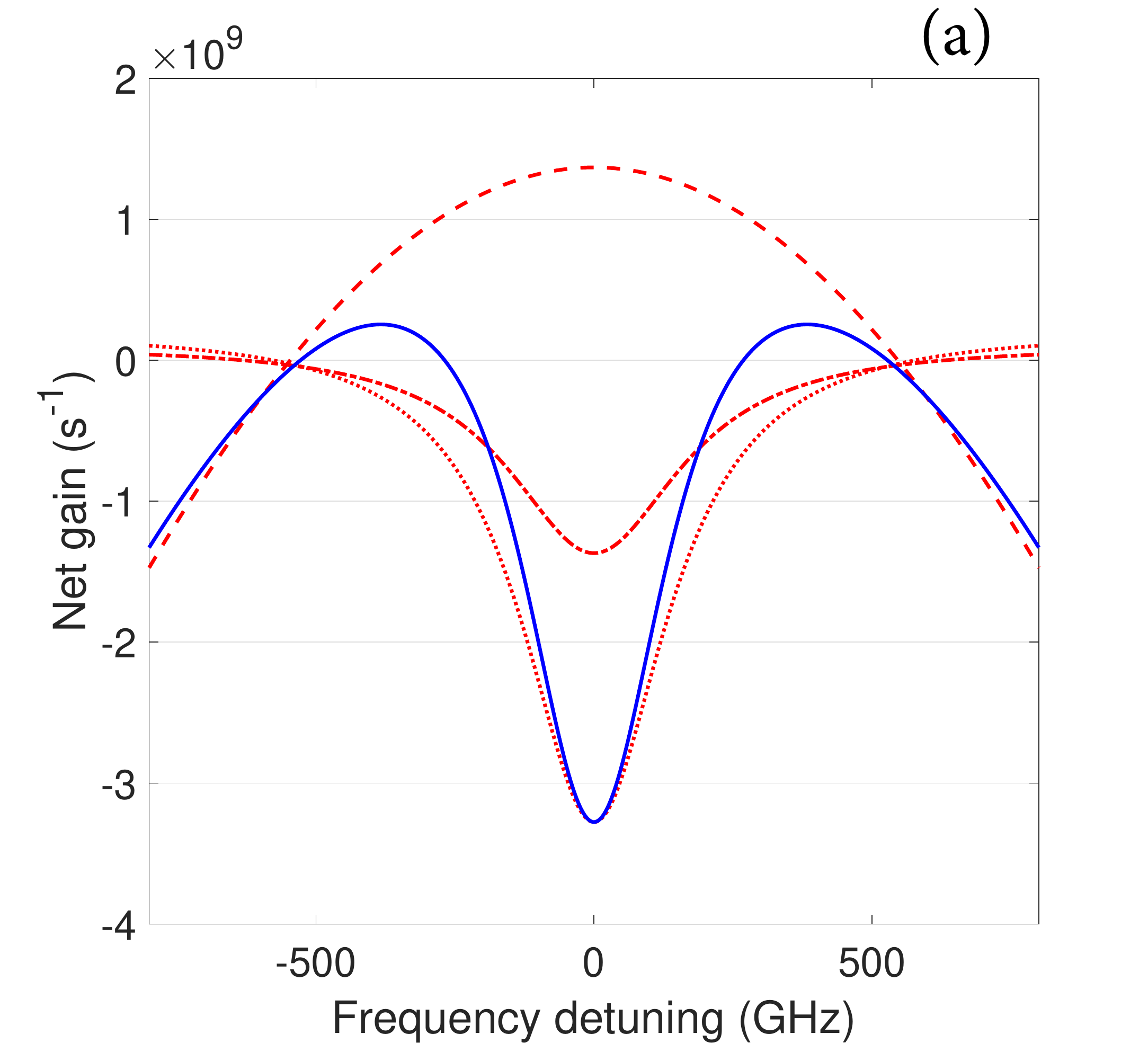}
	\end{subfigure}
	\begin{subfigure}{0.45\textwidth}
		\includegraphics[width=\linewidth]{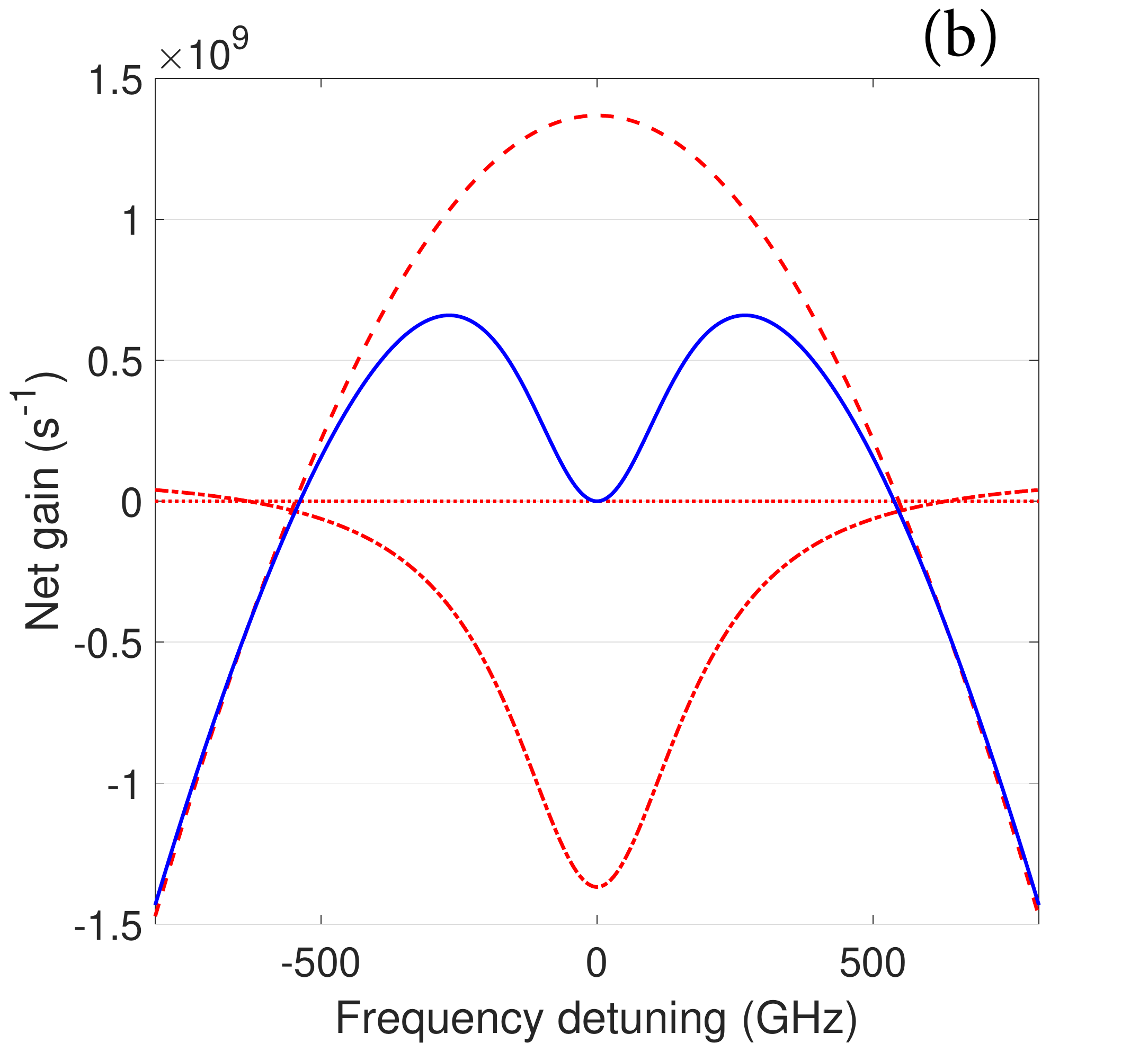}
	\end{subfigure}
	\caption{ The instability gain and contributions from different population terms for (a) $\phi = 0$, and (b) $\phi=\pi$. Solid blue line, red dashed line, red dotted line, and red dash-dot line denote the total gain, $g[\Delta_0]$,  $g[\Delta_{s+}]$, and  $g[\Delta_{2s+}]$, respectively.  Pumping is at $p=1.1$. } .
	\label{Fig:gain_popu}
\end{figure}

To get a better insight into the instability gain, we now analyze the role of population pulsations. With the single mode solution given in \cref{Eq:sol_1mode}, the population pulsations are found to be
\begin{align}
\label{Eq:popu_pulsation}
&\Delta_{s+} = \frac{d^2}{\hbar^2} \frac{\Delta_{th}}{d\omega+i\gamma_1}
\left( \frac{1}{d\omega+i\gamma_2} + \frac{1}{i\gamma_2} \right)
\left( E_0^\ast E_{s+} + E_0 E_{s-}^\ast \right) ,   \nonumber \\
&\Delta_{2s+} = \frac{d^2}{\hbar^2} \frac{\Delta_{th}}{d\omega+i\gamma_g}
\left( \frac{1}{d\omega+i\gamma_2} + \frac{1}{i\gamma_2} \right)
E_0^\ast E_{s+}  ,   \nonumber \\
&\Delta_{2s-} = \frac{d^2}{\hbar^2} \frac{\Delta_{th}}{d\omega+i\gamma_g}
\left( \frac{1}{d\omega+i\gamma_2} + \frac{1}{i\gamma_2} \right)
E_0 E_{s-}^\ast  .
\end{align}
Here we can see that $\Delta_{s+}$ vanishes if $|E_{s+}|=|E_{s-}|$ and $\phi=\pi$, as expected for an FM optical field. However, the pulsations in population grating $\Delta_{2s\pm}$ are always present. Using \cref{Eq:eta_dynamics}, one can see the cross coupling between $E_{s+}$ and $E_{s-}$ originates from $\Delta_{s+}$ only, while the direct coupling is due to both $\Delta_{s+}$ and $\Delta_{2s\pm}$. With the expressions for the population pulsations given in \cref{Eq:popu_pulsation}, the susceptibility of a side mode in an eigenstate can be split into parts corresponding to the population terms in \cref{Eq:eta_dynamics}. For the side mode $E_{s+}$, the susceptibility $\chi_{s+} = \eta_{s+}/E_{s+}$ is written as
\begin{align}
\chi_{s+} = \chi_{s+}[\Delta_0] + \chi_{s+}[\Delta_{s+}] + \chi_{s+}[\Delta_{2s+}] , 
\end{align}
where
\begin{align}
&\chi_{s+}[\Delta_0] = \frac{d}{\hbar} \frac{\Delta_{th}} {d\omega+i\gamma_2}
\left[ 1 + \left(\frac{d|E_0|}{\hbar}\right)^2 \frac{1}{\gamma_g\gamma_2} \right]  ,  \nonumber \\
&\chi_{s+}[\Delta_{s+}] = \frac{d}{2\hbar} \frac{\Delta_{th}} {d\omega+i\gamma_2} 
\left(\frac{d|E_0|}{\hbar}\right)^2 \frac{1}{d\omega+i\gamma_1}
\left( \frac{1}{d\omega+i\gamma_2} + \frac{1}{i\gamma_2} \right)
\left( 1 + \frac{|E_{s-}|}{|E_{s+}|} e^{-i\phi} \right) , \nonumber \\
&\chi_{s+}[\Delta_{2s+}] = \frac{d}{2\hbar} \frac{\Delta_{th}} {d\omega+i\gamma_2} 
\left(\frac{d|E_0|}{\hbar}\right)^2 \frac{1}{d\omega+i\gamma_g}
\left( \frac{1}{d\omega+i\gamma_2} + \frac{1}{i\gamma_2} \right)
 .
\end{align}
Then, the gain can also be separated into parts associated with the population terms. For a dispersionless waveguide refractive index $n(\omega)=n(\omega_0)$, $|E_{s-}|/|E_{s+}| = 1$, then the gains are shown in  \cref{Fig:gain_popu}. Here we have absorbed the loss into the term $g[\Delta_0]$. 


The gain spectrum shows that $g[\Delta_0]$ is positive at low frequency detunings, while $g[\Delta_{2s+}]$ is negative at low frequency detunings. These two contributions result in a gain which is zero at zero frequency detuning, while generally positive at low frequency denunings. The $g[\Delta_{s+}]$ term depends on the phase $\phi$. As we stated earlier, if $\phi=\pi$, there is no population pulsation $\Delta_{s+}$, and hence no contribution to the gain. On the other hand, if $\phi=0$, the gain $g[\Delta_{s+}]$ is again negative for low frequency detunings, so the total gain is also negative. The fact that the gain corresponding to population pulsations is negative can also be understood as a negative feedback to the side modes. In \cref{Fig:gain_popu}, the gain corresponding to population pulsations can be slightly positive at large frequency detunings. This is because we used the assumption $d\omega \ll \{ \gamma_1,~\gamma_g,~\gamma_2 \}$ in obtaining \cref{Eq:eta_dynamics}, so when $d\omega \sim \gamma_1$, the results are not accurate any more. However, $g[\Delta_0]$ is always negative at large frequency detunings, so it does not affect the overall shape of the gain.

\subsection{Effect of group velocity dispersion}
\begin{figure}[htb]
	\centering
	\begin{subfigure}{0.45\textwidth}
		\centering
		\includegraphics[width=\linewidth]{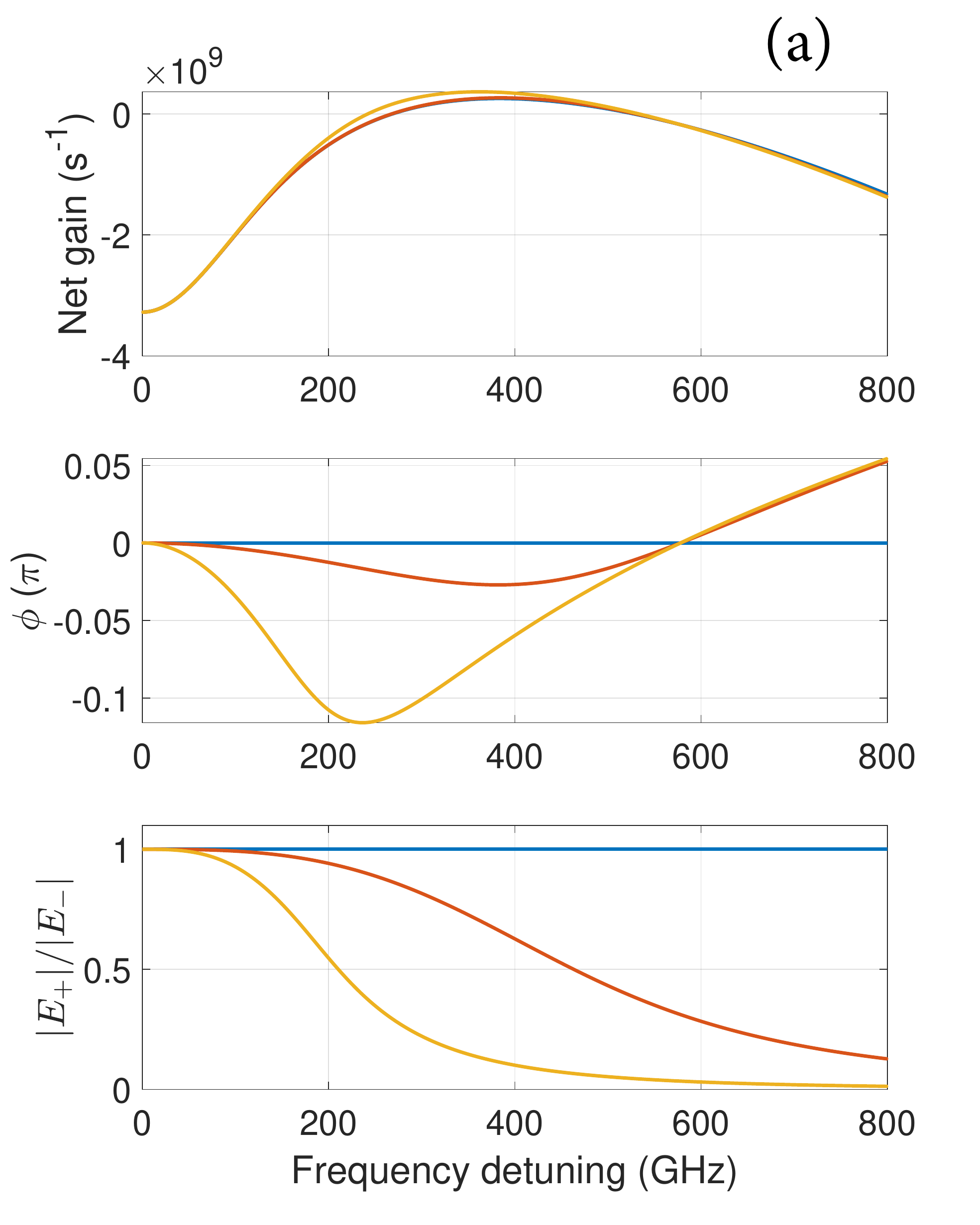}
	\end{subfigure}
	\begin{subfigure}{0.45\textwidth}
		\includegraphics[width=\linewidth]{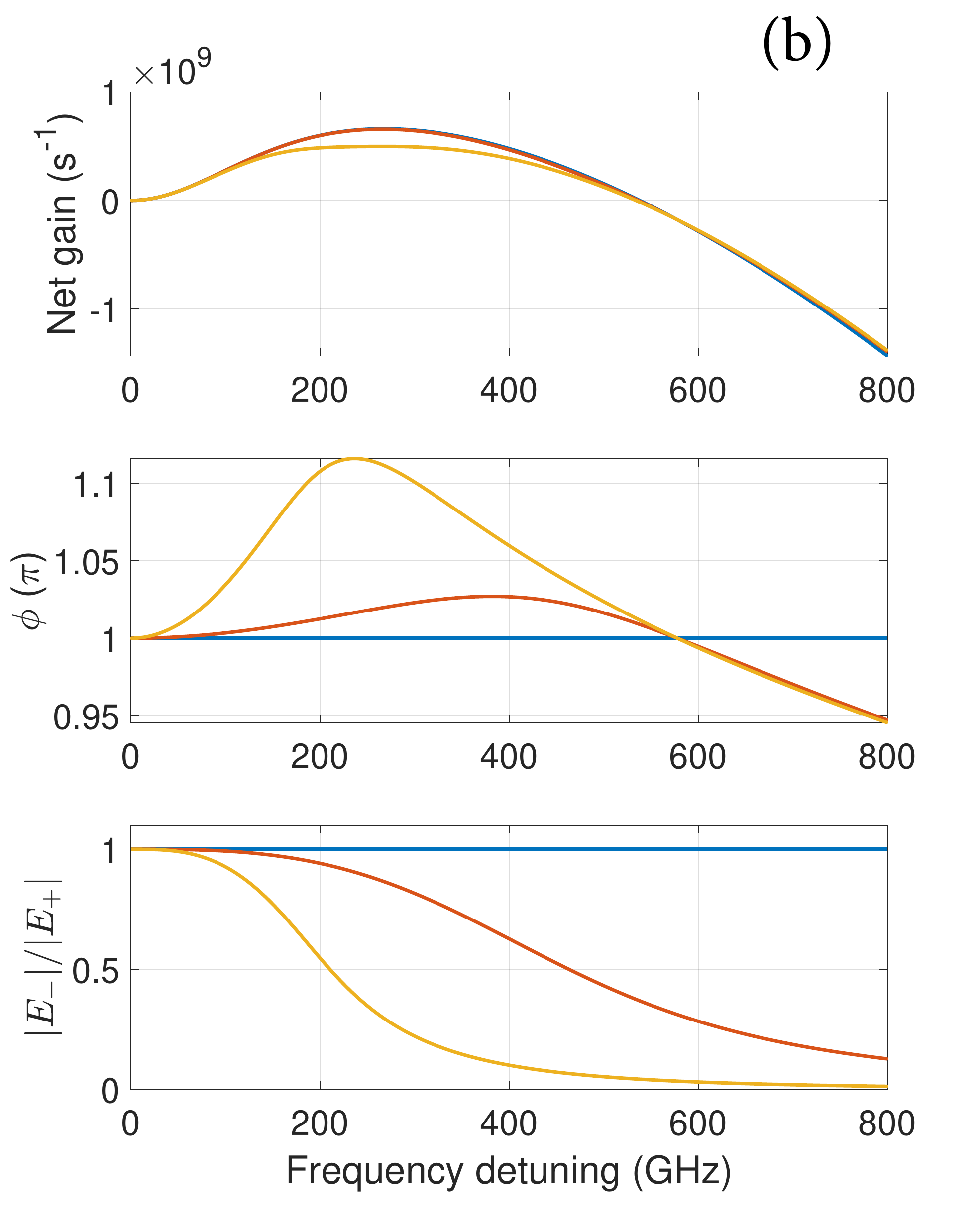}
	\end{subfigure}
		\caption{Dependence of gain spectrum of the weak side modes on GVD around (a) $\phi=0$, and (b) $\phi=\pi$ for the $\beta_{2}$ values of 0 (blue line), $10^3$ (red line) and $10^4$ fs$^2$/mm (yellow line). Other parameters are $T_1$ = 1.0 ps, $T_2$ = 0.05 ps, and $p=1.1$.  }
		\label{Fig:gain_GVD}
\end{figure}

When the waveguide mode has a nonzero GVD, the refractive index of a mode depends on its frequency. In this calculation, we set the following expression for the refractive index:
\begin{equation}
n(\omega) = 3.23 + 0.5 c \beta_{2} (\omega - \omega_{ul}) ,
\end{equation}
where $\beta_{2}$ is the GVD coefficient. Then, we calculate the gain spectrum for different values of $\beta_{2}$, which is shown in \cref{Fig:gain_GVD}, together with the eigenmodes, which are represented by the phase $\phi$ and the amplitude ratio. When $\beta_{2}$ is increased, the gain spectrum itself changes little, while the eigenmode amplitude ratio changes dramatically. Basically the side modes become more asymmetric when the frequency detuning is increased, and the phase $\phi$ of the side modes deviates from $\phi=0$ or $\phi=\pi$ as well. In the presence of GVD, the two side modes with opposite detunings for both frequency and wave vector cannot satisfy the dispersion relations simultaneously without the coupling to the central mode. So, the FWM induced by the strong central mode and gain medium can be viewed as a frequency-pulling mechanism, as shown in our previous work \cite{kazakov2017}.

\section{Linear stability of a three-mode harmonic state}
\label{sec:linear_3mode}

The single-mode instability analysis in previous sections explains how the harmonic state can be favored as the pumping is increased, but does not say anything about its stability. In order to get some insight why the harmonic state can be stable in a QCL, we have to start with the multimode harmonic laser state already present in the cavity and consider its small perturbations. All variables corresponding to a CW harmonic state lasing can be explicitly solved for in frequency domain. Therefore, linearized equations for the small perturbations of the harmonic state can be written explicitly and solved. However, the complexity increases rapidly with the number of strong laser modes due to the nonlinear coupling. To keep the algebra reasonable, we restrict ourselves to  harmonic states with the lowest number of modes, namely three strong modes.

\subsection{Steady-state solution for a three-mode harmonic state}

When three strong modes are present, the two side modes are self-consistently coupled to the central mode and the population inversion. The method for finding such a state is given in Appendix. \ref{sec:appendix_three_mode}. In \cref{Fig:harmonic_state_phi_0_and_pi}, we show the CW mode amplitudes at different frequency detunings as a function of pumping. Here we have assumed that GVD is zero, so the phase difference can be fixed at 0 or $\pi$. When $\phi=0$, we don't find any three-mode solutions for frequency detunings lower than 250 GHz. Also, when $\phi=0$, there exist states where the central mode vanishes, indicating the existence of the parameter space where the two side modes suppress the development of an AM-like central mode. The FM-like harmonic states can happen for small frequency detunings and low pumping levels, consistent with what we already know about the single-mode instability: it has a low threshold and relatively small peak frequencies for FM-like side modes. On the other hand, the AM-like harmonic states are found for higher pumping levels and relatively large frequency detunings, which is also consistent with the single-mode instability for AM-like side modes.

\begin{figure}[htb]
	\centering
	\begin{subfigure}{0.45\textwidth}
		\centering
		\includegraphics[width=\linewidth]{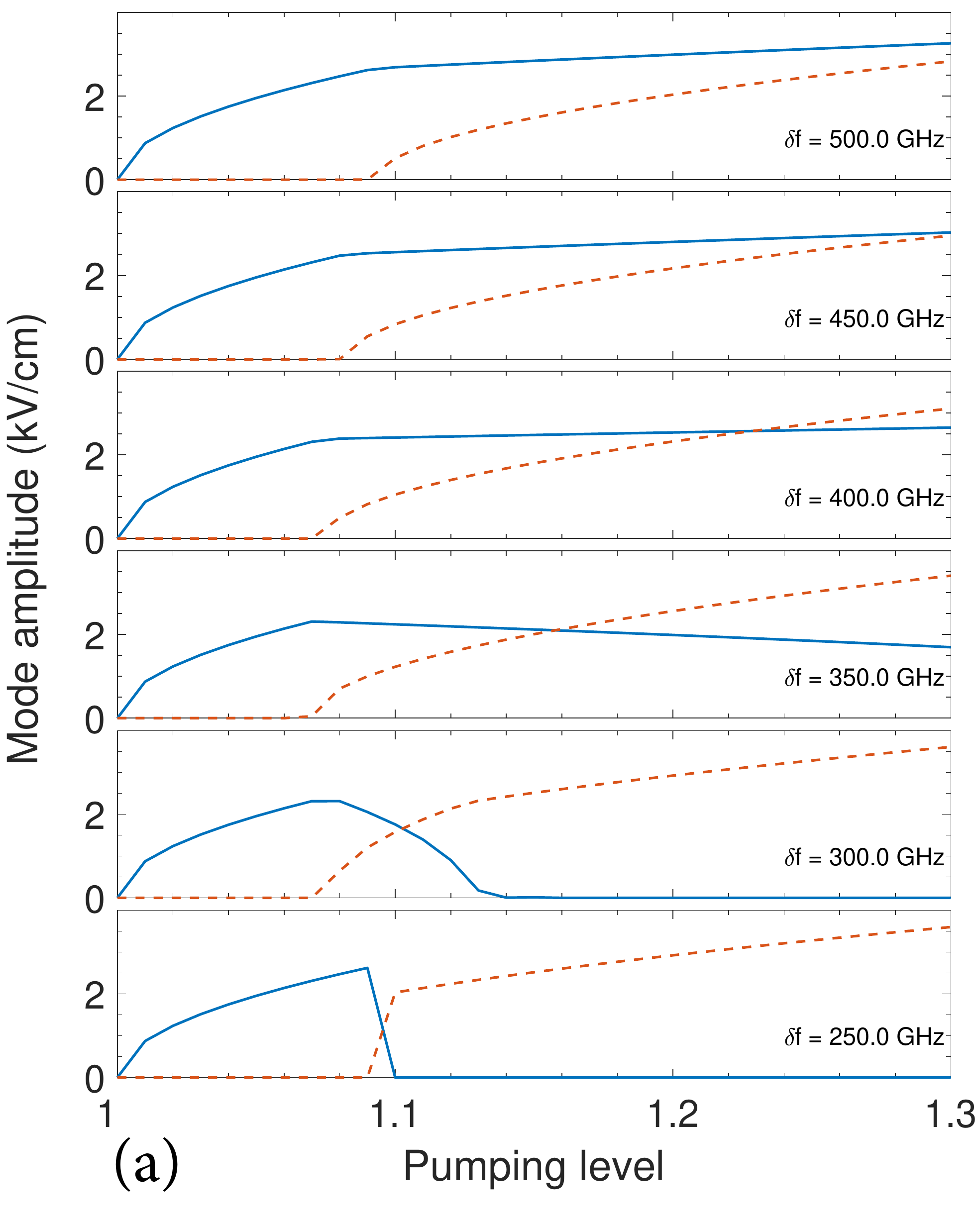}
	\end{subfigure}
	\begin{subfigure}{0.45\textwidth}
		\includegraphics[width=\linewidth]{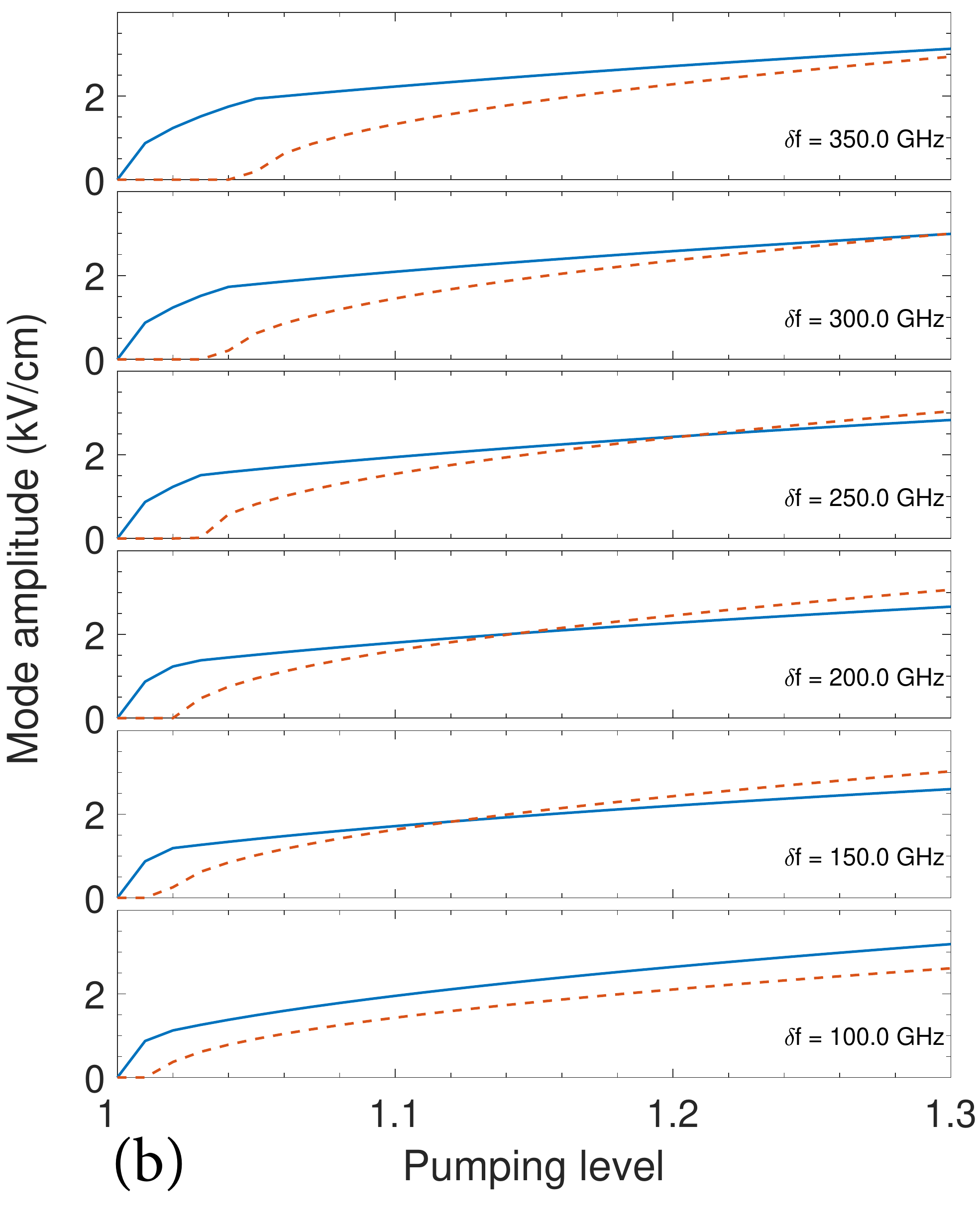}
	\end{subfigure}
	\caption{ The steady-state amplitudes of a three-mode harmonic state as functions of pumping level at different frequency detunings, for (a) $\phi = 0$, and (b) $\phi=\pi$. Blue solid curve is for the central mode, and red dashed line is for the two side modes which have equal amplitudes. }
	\label{Fig:harmonic_state_phi_0_and_pi}
\end{figure}

\subsection{The stability of the harmonic state}
\begin{figure}[htb]
	\centering
		\includegraphics[width=0.9\textwidth]{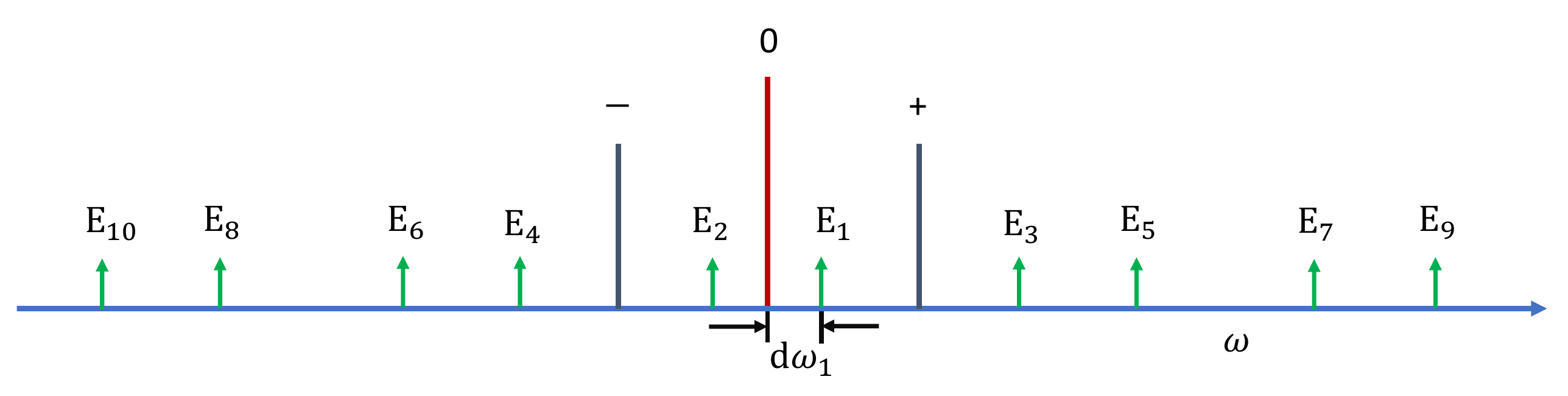}
		\caption{ The scheme for calculating the instability of a three-mode harmonic state. The strong modes in the harmonic state are labeled by $0$, $+$ and $-$, respectively. We consider 10 weak modes, labeled by $E_1$ through $E_{10}$. Mode 0 couples with the following pairs: $\{E_1, E_2\}$, $\{E_3, E_4\}$, $\{E_5, E_6\}$, $\{E_7, E_8\}$, and $\{E_9, E_{10}\}$. Mode $+$ couples with the following pairs: $\{E_1, E_3\}$, $\{E_2, E_5\}$, $\{E_4, E_7\}$, and $\{E_6, E_9\}$.  Mode $-$ couples with the following pairs: $\{E_1, E_6\}$, $\{E_2, E_4\}$, $\{E_3, E_8\}$, and $\{E_5, E_{10}\}$. The frequencies and wave vectors of all modes can be expressed through those of mode $E_1$ }
		\label{Fig:three_mode_stability_scheme}
\end{figure}

The stability of the harmonic state we just found is determined by the gain of other weak modes in the presence of three strong modes. The calculation is similar to that for the instability of a single-mode state, except that weak modes can couple with each other through all of the three strong modes in the harmonic state. The scheme of the coupling is shown in \cref{Fig:three_mode_stability_scheme}. For a given weak mode $E_1$, other modes can be generated due to its FWM interaction with all strong modes. In principle, there can be an infinite number of modes generated. However, the coupling becomes weaker as the two side modes are detuned further from the strong mode, as indicated by the expressions for $\alpha$'s in \cref{Eq:alpha_def}. So, we limit ourselves to 10 weak modes here, and the couplings between them are listed in the caption of \cref{Fig:three_mode_stability_scheme}. After calculating the gain, we need to check if the outermost modes in the resulting eigenstates have relatively small amplitudes. If not, we need to include more weak modes with larger frequency detunings.

The three strong modes are denoted by $0$, $+$ and $-$, and they have frequencies $\omega_0$, $\omega_+$ and $\omega_-$, and wave vectors $k_0$, $k_+$ and $k_-$, respectively. The modes satisfy $\omega_+ + \omega_- = 2\omega_0$ and $k_+ + k_- = 2k_0$. The weak modes are denoted by $E_1$ through $E_{10}$, with frequencies $\omega_1$ through $\omega_{10}$, and wave vectors $k_1$ through $k_{10}$. Among them, we can use $\omega_1$ and $k_1$ as free parameters, and all the others can be determined from them, according to the symmetry with respect to the strong modes.

Similar to the calculation of the instability of a single strong mode, the dynamics of the weak modes is given by 
\begin{align}
\partial_t \left[ E_s \right]
= \left[ B \right] \left[ E_s \right] ,
\end{align}
with
\begin{align}
\left[ E_s \right] =
\begin{bmatrix}
E_1 & E_2^\ast & E_3^\ast & E_4 &  E_5 &  E_6^\ast & E_7^\ast & E_8 & E_9 & E_{10}^\ast
\end{bmatrix}^\mathrm{T} .
\end{align}
The matrix $[B]$ again has the form
\begin{align}
\left[ B \right] = - l_t + \left[ B \right]_{\mathrm{phase}} + \left[ B \right]_{\mathrm{direct}} + \left[ B \right]_{\mathrm{cross}} ,
\end{align}
where the terms have a structure similar  to those in the one-mode stability case, except that now we need to consider all of the couplings. The expressions for $\chi_0(\omega)$ and $\alpha$'s are given in \cref{Eq:chi0_def} and \cref{Eq:alpha_def}. To get them, we need to know the population $\Delta_0$ and the polarizations $\eta_{0,+,-}$ of the three strong modes. To the order needed, they are expressed as
\begin{align}
&\eta_0 = -i \frac{d \Delta_{th} }{\hbar\gamma_2} E_0, \nonumber \\
&\eta_+ = -i \frac{d \Delta_{th} }{\hbar(\gamma_2-i\delta\omega)} E_+, \nonumber \\
&\eta_- = -i \frac{d \Delta_{th} }{\hbar(\gamma_2+i\delta\omega)} E_-, \nonumber \\
&\Delta_0 = \Delta_{th} + (p-1) \Delta_{th} - \frac{2 d^2}{\hbar^2\gamma_1} \Delta_{th} \left[ \frac{1}{\gamma_2}|E_0|^2  + \frac{\gamma_2}{\gamma_2^2+\delta\omega^2} \left(|E_+|^2 + |E_-|^2\right) \right] .
\end{align}

\begin{figure}[htb]
	\centering
	\begin{subfigure}[b]{0.45\textwidth}
	\centering
	\includegraphics[width=\linewidth]{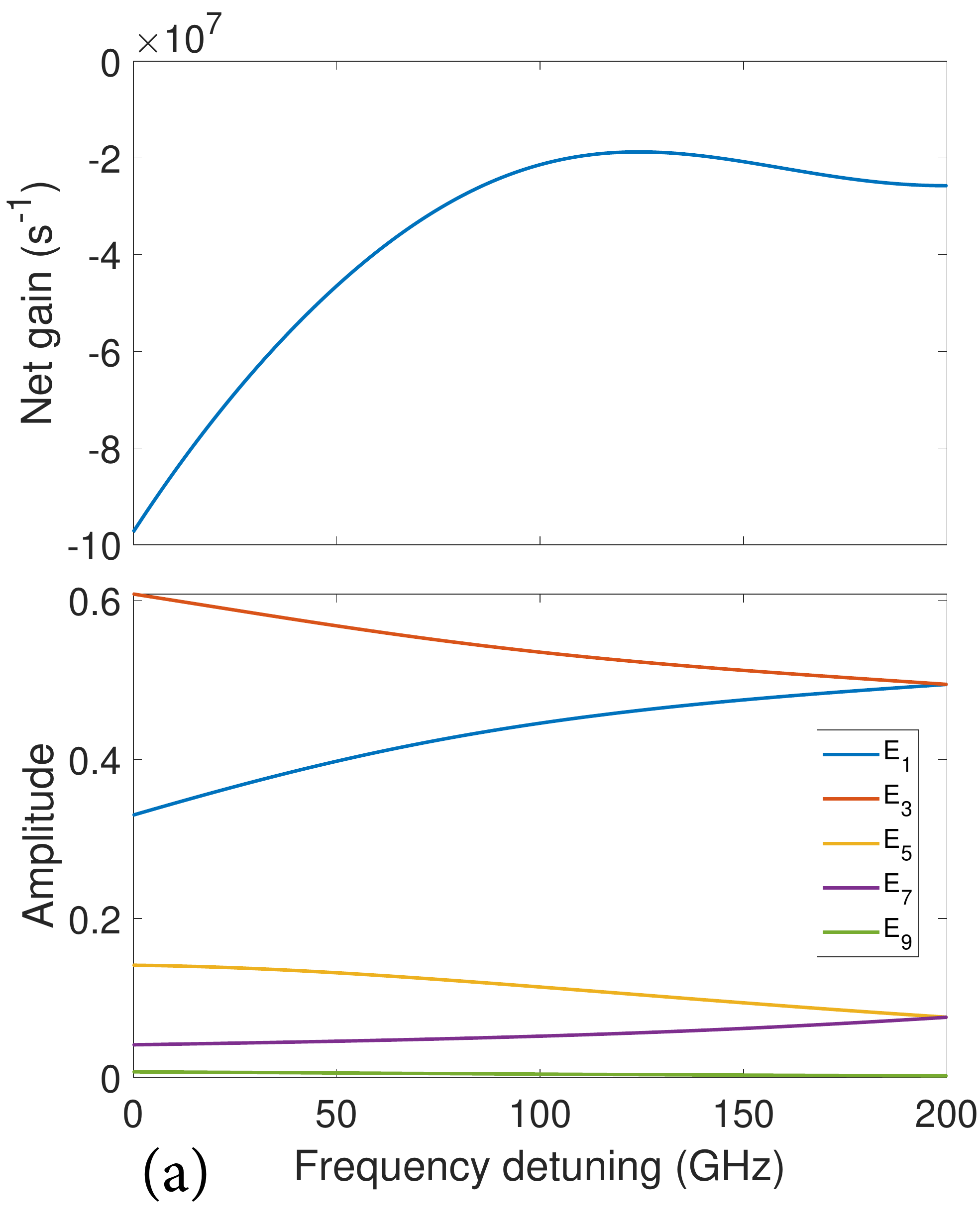}
	\end{subfigure}
	\begin{subfigure}[b]{0.45\textwidth}
		\includegraphics[width=\linewidth]{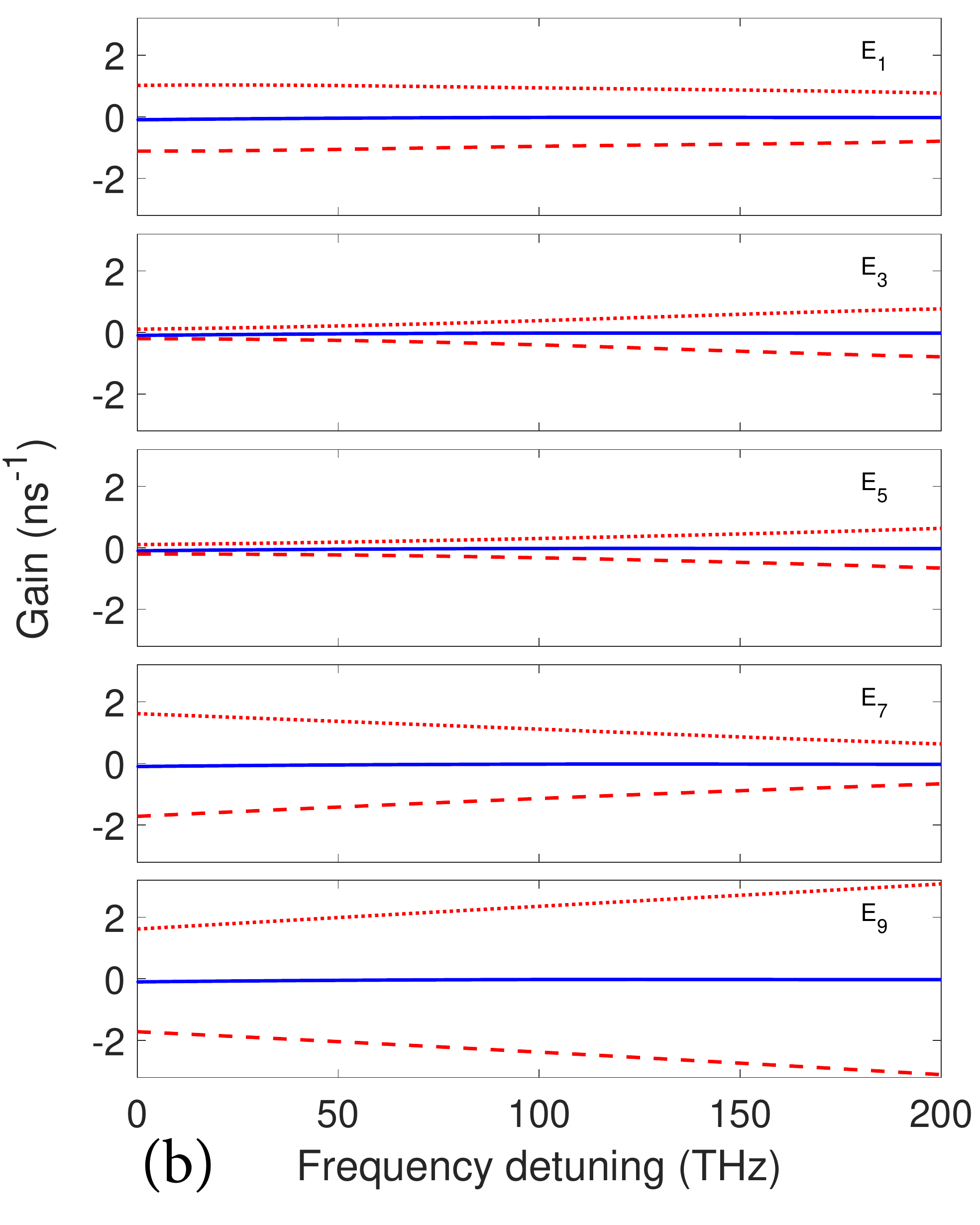}
	\end{subfigure}
	
	\caption{ (a) The instability gain of a three-mode harmonic state (top panel), and the corresponding eigenmodes (bottom panel). (b) The direct gain (red dashed line), cross gain (red dotted line), and the total gain (blue solid line) for each of the side modes. The harmonic state is for the case $\phi=\pi$, $\delta f = 200~\mathrm{GHz}$, and $p=1.10$. Here we only show the side modes $E_1$, $E_3$, $E_5$, $E_7$ and $E_9$, since the other five modes are symmetric to the shown ones with respect to the strong mode 0. }
		\label{Fig:instability_3mode}
\end{figure}

Using the method given above, we calculate the instability gain of the three-mode harmonic state. Particularly we find that the harmonic states with $\phi=0$ all have positive instability gain. However, when $\phi=\pi$, we find several cases where the instability gain is all negative. One such case is shown in \cref{Fig:instability_3mode}. Here we have chosen the eigenstate with the highest instability gain. It can be seen that the weak mode $E_9$ has negligible amplitude at each frequency, which indicates that including 10 modes is enough for the analysis.


The instability gain is a result of the interplay between direct and cross coupling terms. The direct gain and cross-coupling gain for each mode can be defined similar to the single strong mode case. They are shown in \cref{Fig:instability_3mode}(b). The modes $E_3$ and $E_5$ have larger direct gains, while the direct gains for modes $E_7$ and $E_9$ are suppressed. The cross couplings are leading to positive gains, similar to the $\phi=\pi$ case of a single strong mode, see \cref{Fig:gain_direct_and_cross}(b). The direct and cross coupling contributions almost cancel each other, resulting in a common gain shown in \cref{Fig:instability_3mode}(a) (top panel), same for each mode, as they are in an eigenstate. 

The fact that the instability gain is negative at all frequencies indicates that the three-mode harmonic state can be self-supported and stable. From \cref{Fig:instability_3mode}(b), it can be seen that the direct gains are all negative, and the cross gains are positive. They nearly cancel each other and a small negative total gain is left. This indicates that the harmonic state in QCLs results from a delicate interplay of different nonlinear coupling mechanisms, in which phase-sensitive modal interactions play an important role.

\cref{Fig:instability_3mode} is one example of the all-negative instability gain. We have also found other cases of stable harmonic states with $\phi=\pi$ having all-negative gain. They exist at $1.02<p<1.04$ for $\delta f = 100 \,\mathrm{GHz}$, $1.03<p<1.08$ for $\delta f = 150 \,\mathrm{GHz}$, $1.04<p<1.12$ for $\delta f = 200 \,\mathrm{GHz}$, $1.05<p<1.11$ for $\delta f = 250 \,\mathrm{GHz}$, and $1.07<p<1.09$ for $\delta f = 275 \,\mathrm{GHz}$. There is no negative gain region when $\delta f$ is 300 or 350 GHz. This shows that the harmonic states are stable within certain ranges of injection currents and frequency detunings.   


\section{Conclusions}
In conclusion, we have reproduced the experimentally observed harmonic state of mid-IR QCLs in space-time-domain numerical simulations and investigated its formation and stability within frequency-domain analytic linear theory. The numerical simulations  show that a typical harmonic state corresponds to an FM-like optical wave, but strong amplitude modulation may exist at second and possibly higher harmonics of the intermode beat frequency, indicating the direct link from mid-IR lasing to (sub)THz coherent current oscillations in a laser cavity, which can be explored for coherent THz sources and transceivers.  

We studied the instability of a single mode lasing of a QCL and linear stage of the formation of the harmonic state within linearized analytic theory. We showed that the onset of the multimode lasing in a mid-IR QCL is always linked to the formation of a FM-like optical waveform, which is consistent with the numerical simulations and recent experimental studies. The reason the FM gain is larger than the AM gain is because the AM-like side modes lead to population pulsations, which suppress the instability, while the population pulsation is zero for FM-like side modes. The peak gains for both the FM- and AM-like side modes occur at frequencies detuned from the central mode by hundreds of GHz, which is consistent with tens of FSR separation between the modes of a harmonic frequency comb in experiment. 

To study the stability of the harmonic state, we developed an analytic theory describing the growth of small perturbations in the presence a harmonic state with three strong laser modes, where the nonlinear couplings of many weak modes with various frequency detunings are included self-consistently. It is shown that the harmonic states with FM-like phase relations can be stable in a certain range of pumping levels and frequency detunings, i.e.~the gain for any weak sidemode in the presence of strong lasing modes of the harmonic state is negative at all frequencies. The stability is a result of the cancellation between the direct gain and the cross-coupling gain, while the cross couplings also adjust the weight of the direct gain from each weak mode. Our study describes amplitude and phase properties of harmonic frequency combs in mid-IR QCLs and provides insight into the formation and stability of the harmonic states. 

\begin{acknowledgments}

This work has been supported by NSF Award No.~1807336. We thank Tobias Mansuripur, Nikola Opacak, Marco Piccardo, and Benedikt Schwarz  for many helpful discussions. 

\end{acknowledgments}

\appendix

\section{Solution for a three-mode harmonic state}
\label{sec:appendix_three_mode}

Unlike the case of a strong central mode and two weak side modes, the strong side modes in the harmonic state affect the central mode. In fact, all three of them are coupled into a self-consistent steady state. Using $\delta\omega$ to denote the frequency detuning, we can write the fields, coherences, and populations as
\begin{align}
\label{Eq:amplitudes_3modes}
{\cal E} &= E_0 \cos(k_0 z) e^{-i \omega_0 t} + E_+ \cos((k_0+\delta k) z) e^{-i (\omega_0+\delta\omega) t} + E_- \cos((k_0-\delta k) z) e^{-i (\omega_0-\delta\omega) t}   + \mathrm{c.c.} ,  \nonumber \\
\rho_{ul} &= \eta_0 \cos(k_0 z) e^{-i \omega_0 t} + \eta_+ \cos((k_0+\delta k) z) e^{-i (\omega_0+\delta\omega) t} + \eta_- \cos((k_0-\delta k) z) e^{-i (\omega_0-\delta\omega) t}  ,  \nonumber \\
\Delta &= \Delta_0 + \left[ \Delta_{0+} \cos(\delta k z) e^{-i\delta\omega t} + \mathrm{c.c.} \right] + \left[ \Delta_{0++} \cos(2\delta k z) e^{-2i \delta\omega t} + \mathrm{c.c.} \right] \nonumber \\
&+ \Delta_2 \cos(2 k_0 z) + \left[ \Delta_{2+-} \cos(2 k_0 z) e^{-2i\delta\omega t} + \mathrm{c.c.} \right] \nonumber \\
&+ \left[ \Delta_{2+} \cos((2k_0 + \delta k) z) e^{-i\delta\omega t} + \Delta_{2-} \cos((2k_0 - \delta k) z) e^{-i\delta\omega t} + \mathrm{c.c.} \right] \nonumber \\
&+ \Delta_{2++} \cos((2 k_0 + 2\delta k) z)  + \Delta_{2--} \cos((2 k_0 - 2\delta k) z)  .
\end{align}
For reasons already discussed, we will keep the response to the second order in strong fields $|E_0|$, $|E_+|$ and $|E_-|$. The dynamics of coherences and populations is given by
\begin{align}
\label{Eq:polarization}
\partial_t \eta_0 &= \left[ i (\omega_0-\omega_{ul}) - \gamma_2 \right] \eta_0  - i \frac{d}{\hbar} \left( E_0 \Delta_0 + \frac{1}{2} E_0 \Delta_2 + \frac{1}{2} E_+ ( \Delta_{0+}^\ast + \Delta_{2+}^\ast ) + \frac{1}{2} E_- ( \Delta_{0+} + \Delta_{2-} )  \right) ,  \nonumber  \\
\partial_t \eta_+ &= \left( i (\omega_0+\delta\omega-\omega_{ul}) - \gamma_2 \right) \eta_+ \nonumber \\ 
&- i \frac{d}{\hbar} \left( \frac{1}{2} E_0 \Delta_{0+} + \frac{1}{2} E_0 \Delta_{2+} + E_+ \Delta_0 + \frac{1}{2} E_+ \Delta_{2++} + \frac{1}{2} E_- \Delta_{0++} + \frac{1}{2} E_- \Delta_{2+-} \right)  ,  \nonumber  \\
\partial_t \eta_- &= \left( i (\omega_0-\delta\omega-\omega_{ul}) - \gamma_2 \right) \eta_- \nonumber \\ 
&- i \frac{d}{\hbar} \left( \frac{1}{2} E_0 \Delta_{0+}^\ast + \frac{1}{2} E_0 \Delta_{2-}^\ast + \frac{1}{2} E_+ \Delta_{0++}^\ast + \frac{1}{2} E_+ \Delta_{2+-}^\ast + E_- \Delta_0 + \frac{1}{2} E_- \Delta_{2--} \right) ,
\end{align}
and 
\begin{align}
\label{Eq:population}
&\partial_t \Delta_0 = - \gamma_1 (\Delta_0-\Delta_p) - i \frac{d}{\hbar} \left(  E_0^\ast \eta_0 - E_0 \eta_0^\ast + E_+^\ast \eta_+ - E_+ \eta_+^\ast + E_-^\ast \eta_- - E_- \eta_-^\ast \right) , \nonumber \\
&\partial_t \Delta_2 = - \gamma_g \Delta_2 - i \frac{d}{\hbar} \left( E_0^\ast \eta_0 - E_0 \eta_0^\ast \right) , \nonumber \\
&\partial_t \Delta_{0+} = \left( i\delta\omega - \gamma_1 \right) \Delta_{0+} - i \frac{d}{\hbar} \left( E_0^\ast \eta_+ - E_0 \eta_-^\ast - E_+ \eta_0^\ast + E_-^\ast \eta_0 \right) , \nonumber \\
&\partial_t \Delta_{0++} = \left( 2i\delta\omega - \gamma_1 \right) \Delta_{0++} - i \frac{d}{\hbar} \left( - E_+ \eta_-^\ast + E_-^\ast \eta_+ \right) , \nonumber \\
&\partial_t \Delta_{2+} = \left( i\delta\omega - \gamma_g \right) \Delta_{2+} - i \frac{d}{\hbar} \left( E_0^\ast \eta_+ - E_+ \eta_0^\ast \right) ,  \nonumber \\
&\partial_t \Delta_{2-} = \left( i\delta\omega - \gamma_g \right) \Delta_{2-} - i \frac{d}{\hbar} \left( - E_0 \eta_-^\ast + E_-^\ast \eta_0 \right)  \nonumber \\
&\partial_t \Delta_{2++} = - \gamma_g \Delta_{2++} - i \frac{d}{\hbar} \left( E_+^\ast \eta_+ - E_+ \eta_+^\ast \right) ,  \nonumber \\
&\partial_t \Delta_{2--} = - \gamma_g \Delta_{2--} - i \frac{d}{\hbar} \left( E_-^\ast \eta_- - E_- \eta_-^\ast \right) ,  \nonumber \\
&\partial_t \Delta_{2+-} = \left( 2i \delta\omega - \gamma_g \right) \Delta_{2+-} - i \frac{d}{\hbar} \left( - E_+ \eta_-^\ast + E_-^\ast \eta_+ \right) .
\end{align}
Putting the time derivatives to be zero in Eqs.~(\ref{Eq:polarization}, \ref{Eq:population}), we are able to find the expressions for the coherences. We assume that $\omega_0=\omega_{ul}$ for simplicity. The expressions for the coherences are given below:
\begin{align}
\eta_0 &= -i \frac{d\Delta_{th}}{2\hbar\gamma_2} \left[ 2 p E_0 - \frac{2 d^2 |E_0|^2 E_0}{\hbar^2 \gamma_2} \left( \frac{2}{\gamma_1} + \frac{1}{\gamma_g} \right) \right. \nonumber \\
&- \left. \frac{d^2|E_-|^2 E_0}{\hbar^2} \left( \frac{4\gamma_2}{\gamma_1(\gamma_2^2+\delta\omega^2)}  + \left( \frac{1}{\gamma_2}+\frac{1}{\gamma_2-i\delta\omega} \right) \left( \frac{1}{\gamma_1-i\delta\omega}+\frac{1}{\gamma_g-i\delta\omega} \right) \right) \right. \nonumber \\
&- \left. \frac{d^2|E_+|^2 E_0}{\hbar^2} \left( \frac{4\gamma_2}{\gamma_1(\gamma_2^2+\delta\omega^2)}  + \left( \frac{1}{\gamma_2}+\frac{1}{\gamma_2+i\delta\omega} \right) \left( \frac{1}{\gamma_1+i\delta\omega}+\frac{1}{\gamma_g+i\delta\omega} \right) \right) \right. \nonumber \\
&- \left. \frac{2 d^2 E_0^\ast E_+ E_-}{\hbar^2\gamma_2} \frac{\gamma_1\delta\omega^2+2\gamma_1\gamma_2^2-\gamma_2\delta\omega^2}{(\gamma_1^2+\delta\omega^2)(\gamma_2^2+\delta\omega^2)}  \right] , \\
\eta_+ &= -i \frac{d\Delta_{th}}{2\hbar(\gamma_2-i\delta\omega)} 
\left[
2 p E_+ \phantom{\frac{d^2 |E_0|^2 E_+}{\hbar^2}} \right. \nonumber \\
&- \frac{d^2 |E_0|^2 E_+}{\hbar^2} \frac{2\gamma_1^2\gamma_2+6\gamma_1\gamma_2\gamma_g - i\delta\omega(\gamma_1^2+8\gamma_1\gamma_2+5\gamma_1\gamma_g+4\gamma_2\gamma_g) - \delta\omega^2(6\gamma_1+4\gamma_2+4\gamma_g) + 4i\delta\omega^3}{\gamma_1\gamma_2(\gamma_1-i\delta\omega)(\gamma_g-i\delta\omega)(\gamma_2-i\delta\omega)} \nonumber \\
& - \frac{2 d^2 |E_+|^2 E_+}{\hbar^2} \frac{\gamma_2(\gamma_1+2\gamma_g)}{\gamma_1\gamma_g(\gamma_2^2+\delta\omega^2)}
\nonumber \\
& - \frac{2 d^2 |E_-|^2 E_+}{\hbar^2} \left( \frac{2\gamma_2}{\gamma_1(\gamma_2^2+\delta\omega^2)}  + \frac{\gamma_1+\gamma_g-4i\delta\omega}{(\gamma_1-2i\delta\omega)(\gamma_g-2i\delta\omega)(\gamma_2-i\delta\omega)} \right)  \nonumber \\
&- \left. \frac{d^2 E_-^\ast E_0^2}{\hbar^2} \frac{2\gamma_2-i\delta\omega}{\gamma_2(\gamma_1-i\delta\omega)(\gamma_2-i\delta\omega)} 
\right] ,   \\
\eta_- &= -i \frac{d\Delta_{th}}{2\hbar(\gamma_2+i\delta\omega)} 
\left[
2 p E_- \phantom{\frac{d^2 |E_0|^2 E_+}{\hbar^2}} \right. \nonumber \\
&- \frac{d^2 |E_0|^2 E_-}{\hbar^2} \frac{2\gamma_1^2\gamma_2+6\gamma_1\gamma_2\gamma_g + i\delta\omega(\gamma_1^2+8\gamma_1\gamma_2+5\gamma_1\gamma_g+4\gamma_2\gamma_g) - \delta\omega^2(6\gamma_1+4\gamma_2+4\gamma_g) - 4i\delta\omega^3}{\gamma_1\gamma_2(\gamma_1+i\delta\omega)(\gamma_g+i\delta\omega)(\gamma_2+i\delta\omega)} \nonumber \\
& - \frac{2 d^2 |E_+|^2 E_-}{\hbar^2} \left( \frac{2\gamma_2}{\gamma_1(\gamma_2^2+\delta\omega^2)}  + \frac{\gamma_1+\gamma_g+4i\delta\omega}{(\gamma_1+2i\delta\omega)(\gamma_g+2i\delta\omega)(\gamma_2+i\delta\omega)} \right)  \nonumber \\
& - \frac{2 d^2 |E_-|^2 E_-}{\hbar^2} \frac{\gamma_2(\gamma_1+2\gamma_g)}{\gamma_1\gamma_g(\gamma_2^2+\delta\omega^2)}
\nonumber \\
&- \left. \frac{d^2 E_+^\ast E_0^2}{\hbar^2} \frac{2\gamma_2+i\delta\omega}{\gamma_2(\gamma_1+i\delta\omega)(\gamma_2+i\delta\omega)} 
\right] .
\end{align}
Substituting the coherences into the wave equation \cref{Eq:MWeq_simplified} for each of the three modes, we get a set of three coupled differential equations. The equations can be run numerically to get the steady state. During this process, the detuning of the wave vectors is tuned to ensure that  the steady state can be reached.

\bibliography{QCL_harmonic_ref}

\end{document}